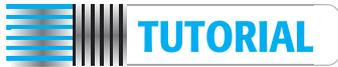

## TUTORIAL

# A Tutorial on Trusted and Untrusted Non-3GPP Accesses in 5G Systems—First Steps Toward a Unified Communications Infrastructure

**MARIO TEIXEIRA LEMES**[1,2], **ANTONIO MARCOS ALBERTI**[3], **CRISTIANO BONATO BOTH**[4], **ANTONIO CARLOS DE OLIVEIRA JÚNIOR**[2,5], (Member, IEEE), AND **KLEBER VIEIRA CARDOSO**[2], (Member, IEEE)

[1]Instituto Federal de Educação, Ciência e Tecnologia de Goiás, Formosa 73813-816, Brazil
[2]Instituto de Informática, Universidade Federal de Goiás, Goiânia 74690-900, Brazil
[3]Instituto Nacional de Telecomunicações, Santa Rita do Sapucaí 37540-000, Brazil
[4]Applied Computing Graduate Program, Universidade do Vale do Rio dos Sinos, São Leopoldo 93022-750, Brazil
[5]Fraunhofer AICOS, 4200-135 Porto, Portugal

Corresponding author: Mario Teixeira Lemes (mario.lemes@ifg.edu.br)

This work was supported in part by the National Education and Research Network [Rede Nacional de Ensino e Pesquisa (RNP)] with resources from Ministry of Science, Technology, Innovations, and Communications [Ministério da Ciência, Tecnologia, Inovações e Comunicações (MCTIC)], under the Brazil 6G Project of the Radiocommunication Reference Center (Centro de Referência em Radiocomunicações—CRR) of the National Institute of Telecommunications (Instituto Nacional de Telecomunicações—Inatel), Brazil, under Grant 01245.010604/2020-14.

**ABSTRACT** fifth-generation (5G) systems are designed to enable convergent access-agnostic service availability. This means that 5G services will be available over 5G New Radio air interface and also through other non-Third Generation Partnership Project (3GPP) access networks, e.g., IEEE 802.11 (Wi-Fi). 3GPP has recently published the Release 16 that includes trusted non-3GPP access network concept and wireless wireline convergence. The main goal of this tutorial is to present an overview of access to 5G core via non-3GPP access networks specified by 3GPP until Release 16 (i.e., untrusted, trusted, and wireline access). The tutorial describes convergence aspects of a 5G system and these non-3GPP access networks, such as the authentication and authorization procedures and the data session establishment from the point of view of the protocol stack and exchanged messages among the network functions. In order to illustrate several concepts and part of 3GPP specification, we present a basic but fully operational implementation of untrusted non-3GPP access using WLAN. We perform experiments illustrating how a Wi-Fi user is authorized in a 5G core, establishing user plane connectivity to a data network. Moreover, we evaluate the performance of this access in terms of time consumed, the number of messages, and protocol overhead to established data sessions.

## I. INTRODUCTION

The convergence of 5G systems and non-3GPP access networks may have a deep impact in the Information and Communication Technology (ICT) context. 5G Core Network (5GCN) is highly flexible due to the adoption of Service-Based Architecture (SBA), network slicing, Software-Defined Networking (SDN), among other modern paradigms [1]. In addition to the Next-Generation Radio Access Network (NG-RAN) based on 5G New Radio

(5G-NR), 5GCN was designed for integrating multiple access networks, e.g., Long Term Evolution (LTE)/4G and Wireless Local Area Network (WLAN), simply and efficiently. Integrating 3GPP and non-3GPP access networks is fundamental in the adoption of 5G Non-Public Networks (NPNs) for verticals domains that employ heterogeneous access networks, e.g., Wi-Fi is ubiquitous in this context. Intelligent integration between 3GPP and non-3GPP access networks, with effective solutions to relieve data congestion, address capacity, and coverage issues, is crucial to address the new use cases resulting from the explosive growth of Internet of Things (IoT) devices and industrial communication [2].

The associate editor coordinating the review of this manuscript and approving it for publication was Zhenzhou Tang.

Furthermore, this integration enables end-devices accessing through non-3GPP access network and with no 5G capabilities, e.g., legacy and IoT devices benefit from 5G scenarios: (i) Enhanced Mobile Broadband (eMBB) for greater bandwidth, (ii) Massive Machine Type Communications (mMTC) for high connection density, and (iii) Ultra-Reliable Low Latency Communications (URLLC) for end-to-end latency reduction. The seamless interworking between 5G access and other industry access technologies minimizes operational costs to widespread adoption of 5G networks, especially in the initial phases [3].

Historically, there has been a desire to unify heterogeneous access networks to mobile cellular technologies. Salkintzis [4] highlights the strong need to integrate WLAN with 3G/4G to support the development of hybrid mobile data networks, indicating the importance of WLAN access network technology in the context of mobile cellular networks. The author provides six different architectures with heterogeneous requirements that can support interworking between WLAN and 3G networks, showing the high complexity of this integration process. Ferrus et al. [5] propose mechanisms to achieve seamless integration in heterogeneous wireless networks, providing a generic interworking scenario with various levels of integration. The authors introduced new hardware elements in the definition of generic access in the 3GPP network. The legacy of the monolithic structure offered by 3G and 4G cores in the mobile cellular network and dedicated hardware solutions increase the challenge of integrating network access. However, in contrast to the previous mobile communication generations, 5G revisits this attempt to integrate and unify different access networks with a new software perspective [6].

Releases of Technical Specification (TS) by 3GPP, for instance [7], recognize the relevance of other consolidated access network technologies, e.g., Wi-Fi, providing flexibility in the access of non-3GPP networks. Mainly from Release 16 (Rel-16) [8] with the addition of trusted non-3GPP access network and wireline access support, the same 5GCN could be used to provide services to a wide range of wireless and wireline access technologies, enabling integration and convergence between new and legacy networks. The wireless wireline convergence facilitates the Mobile Network Operators (MNOs)'s management tasks by providing a single-core network. On the one hand, the standard flexibility enables the integration of heterogeneous access networks suitable for vertical domains, especially for developing new applications for IoT and Industry 4.0 [9]. On the other hand, lack of some information creates further open issues, e.g., the standard does not specify how to establish trust in trusted non-3GPP networks, and there are missing details about the connection of end-devices with no 5G capabilities. However, these issues in the standard of non-3GPP access are not restricted and they bring business opportunities for innovation and development of differentiated solutions to improve the user experience in 5G networks.

Kunz and Salkintzis [10] describe that Release 15 (Rel-15) of 3GPP focuses on security aspects of untrusted non-3GPP access networks and the leading procedures related to registration and authorization. However, the Packet Data Unit (PDU) session establishment permitting the creation of data sessions between User Equipment (UE) and 5G is not discussed. The authors provide an outlook of Rel-16 with trusted non-3GPP access option, but in a similar way to untrusted non-3GPP access, they do not provide related details about connection establishment for transport data sessions among end-devices and 5GCN. Mur et al. [11] investigate how private 5G networks should evolve the non-3GPP access networks, especially in Rel-16 and beyond. These networks should deliver a 5G solution to verticals that integrate 5G-NR, Wi-Fi, and Light Fidelity (Li-Fi), all operating through Artificial Intelligence-based autonomic networking. The integration between 5G and non-3GPP access networks, e.g., Wi-Fi and Li-Fi, is achieved by a link-layer SDN controller that supports seamless mobility between the communication technologies. The intelligence and autonomic features refer to: (i) high-level intent language to operate the network and (ii) Machine Learning (ML) models to support the operation of Network Functions (NFs), i.e., to make predictions or detect anomalies.

The integration between wireless non-3GPP networks and 5G in verticals enables peak data rates, improved area capacity, low delay, and localization enhancements. The authors do not consider wireline access technologies and the wireless wireline convergence is missing in the integration proposal to support a single 5GCN. The literature still lacks tutorial work that summarizes the relevant information about non-3GPP access that was introduced in 3GPP Rel-15 and Rel-16. The relevance of non-3GPP technologies is already identified in the design of future networks, for example, Tataria et al. [12] introduce *superconvergence* of non-3GPP wireless and wireline access technologies as a 6G network design principle.

This article focuses on providing a tutorial on 5GCN access via non-3GPP access networks, considering untrusted, trusted, and wireline, as specified in the last releases of 3GPP TS. To properly understand how non-3GPP access networks are authorized in the 5GCN and provide data services for end-devices, e.g., UE, it is essential to learn about: (i) types of non-3GPP access networks, (ii) main components employed for authentication and authorization procedures, and (iii) data sessions related procedures. Furthermore, the non-3GPP access networks establish different types of trust in communication and support specified contexts, i.e., related to roaming services. This tutorial also enables and may accelerate the development of new applications, consequently expanding the 5G services, especially in vertical domains. More specifically, our main contributions are:

- To provide a comprehensive overview on types of non-3GPP access networks, focusing on the different

possibilities of connection, e.g., wireless and wireline technologies.
- To detail the authorization and authentication procedures, as well as the establishment of user data sessions, focusing on the main messages exchanged and the protocols involved among 5GCN components, non-3GPP access networks, and end-devices.
- To show the connection between Wi-Fi access network and 5GCN as Proof-of-Concept (PoC) to consolidate tutorial information on the untrusted non-3GPP access network.

The sections of this work are organized as follows. In Section II, we introduce the scope considered in this tutorial, providing an overview of the 5G system and the non-3GPP access networks. To facilitate the identification and the possibilities of non-3GPP access networks, we present a comparison of untrusted, trusted, and wireline access networks, leaving the specific details of each access network to individual sections. The untrusted and trusted networks are detailed in Section III and Section IV, respectively. In Section V, we provide details about 5GCN access using wireline networks. To illustrate the procedures related to untrusted networks, we present a non-3GPP access Wi-Fi use case in Section VI, as a simple PoC of the integration between 3GPP and non-3GPP access networks. Furthermore, we evaluate the PoC performance considering time consumed, number of messages, and protocol overhead to establish data session metrics. Final remarks and additional future research considerations are stated in Section VII.

## II. OVERVIEW OF 5G SYSTEM AND NON-3GPP ACCESS NETWORKS
In this section, we present basic concepts about 5GCN and access networks in general, as specified in 3GPP Rel-15 and Rel-16. We also briefly introduce the three non-3GPP access networks, i.e., untrusted, trusted, and wireline, and compare some of their relevant characteristics, such as communication security, UE Non-Access Stratum (NAS) signaling, and roaming support.

### A. 5G SYSTEM
Basically, the 5GS consists of Radio Access Network (RAN) and Core Network (CN). The introduction of SBA in 5GCN is a significant change in comparison with 4G/Evolved Packet Core (EPC) and the previous generations of mobile communications. From 5GCN, the core can be visualized and described differently. From one point of view, 5GCN is a collection of NFs that provides the expected core functionalities. Each NF exposes its services to other NFs, acting as a service producer. As consumers, an NF can use the services offered by other NFs. Such ability exposes and makes services available, characterizing the so-called Service-Based Interface (SBI) [13]. From another point of view, 5GCN has several point-to-point communications, known as *reference points*. This representation describes how NFs interact with each other. A producer-consumer model is a framework that defines the interaction among NFs [14].

The services offered by an NF in 5GCN can be selected, for instance, by UE for core access and mobility management. Any UE needs to establish transport sessions for data transfer and maintain continuous communication with 5GCN for several control and management tasks. NAS protocol [15] is adopted to control message exchanging between UE and 5GCN. Figure 1 illustrates a 5GS SBA reference architecture and how SBA delivers services as a collection of NFs using the Control and User Plane Separation (CUPS) concept [16], [17]. The following NFs are shown in the figure: (i) Access and Mobility Function (AMF), (ii) Session Management Function (SMF), (iii) User Plane Function (UPF), (iv) Authentication Server Function (AUSF), (v) Network Slice Selection Function (NSSF), (vi) Network Exposure Function (NEF), (vii) Network Repository Function (NRF), (viii) Policy Control Function (PCF), (ix) Unified Data Management (UDM), and (x) Application Function (AF). This list of NFs is far from being comprehensive, not including, for example, NFs related to non-3GPP access that we will introduce later.

In the Control Plane (CP), AMF is in charge of mobility management along with the possible handovers of a user. SMF is responsible for maintaining the existing session. AUSF and UDM are standardized to create and manage authentication keys to perform UE authentication and authorization. NSSF, NEF, NRF, PCF, and AF also belong to CP and are important in many control and management tasks, but they are out of scope in this article. In the User Plane (UP), UPF forwards the traffic between UEs and Data Network (DN) [18]. Furthermore, SMF instructs UPF to create packet detection and forwarding rules. To consume services provided by an MNO, UE shall connect over the air interface to RAN, i.e., the Next-Generation Node B (gNodeB), and then requests NAS signaling processing at AMF and PDU session establishment. NFs in SBA communicate one another over SBI using the Hypertext Transfer Protocol (HTTP) [19] and the Transport Layer Security (TLS) for security connection [20], or through the reference points using transport and application layer-specific protocols.

An NF exposes and consumes services via reference points using the producer-consumer model. For instance, NAS signaling between UE and AMF is performed via N1. The N2 interface is the point-to-point communication between gNodeB and AMF for transferring of session management messages. The N3 interface between gNodeB and UPF is used for exchanging packets in UP, whereas N11 is used for AMF and SMF interactions. The N4 interface is employed by SMF for sending rules of packet detection and forwarding to UPF. Finally, the N6 interface connects UPF and DN, which is commonly the Internet.

Most of the time, an information producing NF in a Public Land Mobile Network (PLMN) offers mobile services for UEs connected to a 5G-NR and/or non-3GPP access network inside the Home Public Land Mobile Network (HPLMN), i.e., UE consumes services in the same local that the subscriber profile is configured. However, an NF may also offer services for UEs outside HPLMN, i.e., when they are roaming. As in previous generations, roaming in 5G allows a UE to employ mobile services outside its coverage area or in a Visited Public Land Mobile Network (VPLMN). In this article, we adopt the term *home network* as equivalent to HPLMN and *visited network* as an alternative to any external network (VPLMN) that provides mobile services to a UE outside its home network.

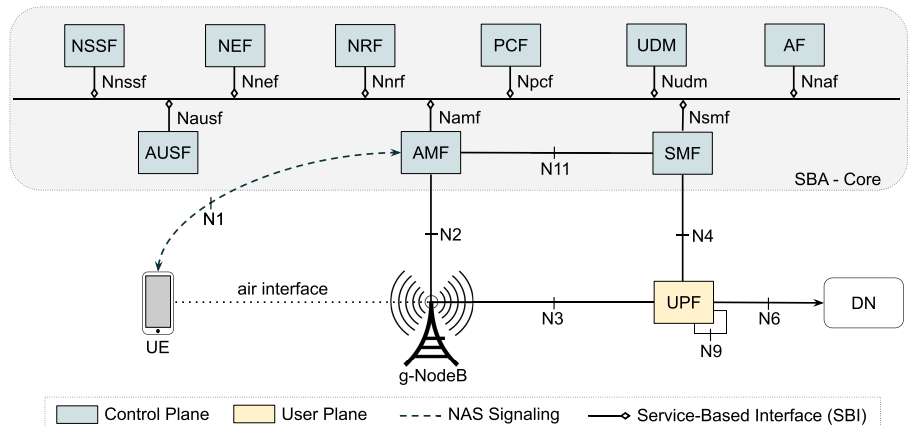

**FIGURE 1.** 5GS SBA reference architecture.

Figure 2 illustrates a simplified roaming 5GS architecture with Local Breakout (LBO) and Home Routed (HR) scenarios, which can be applied to untrusted, trusted and wireline non-3GPP access networks. In the LBO roaming case, the user data traffic, e.g., resulting from PDU session establishment initiated by UE, is routed from the visited network to DN. UE connects to gNodeB in the visited network and consumes NF services, such as AMF for CP and SMF for UP functions from this network. Only the authentication procedure and the subscription handling are performed in the home network [21]. In HR roaming, the signaling data related to the authentication procedure and the visited network data traffic resulting from PDU session establishment are routed to DN from the home network. HR roaming provides additional control to MNO, such as accounting and billing information. However, this approach brings extra complexity and delay to communication. For example, the roaming user experiences increased delay, as discussed in [22] and [23]. To protect the communication between the home network and visited network, a Security Edge Protection Proxy (SEPP), or simply *proxy*, is used in LBO and HR roaming scenarios. This proxy provides message filtering and policing for inter-PLMN control plane interfaces, hiding network topology from other vendors [24].

### B. SUPPORT FOR NON-3GPP ACCESS NETWORKS

To support the connectivity of UE via a non-3GPP access network, the following types of access are specified: (i) untrusted, (ii) trusted, and (iii) wireline [7]. In the following, we briefly introduce each type of these access networks, which is later described in detail.

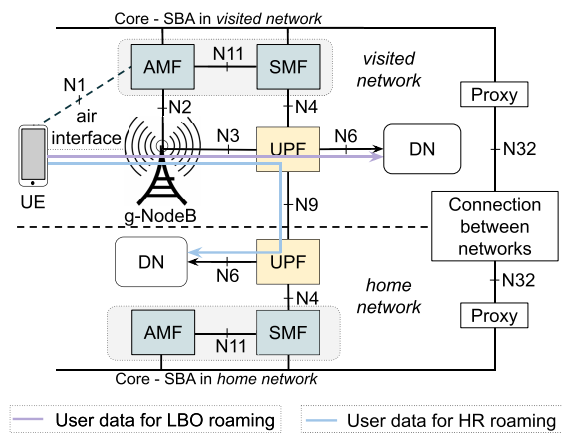

**FIGURE 2.** Roaming 5GS architecture - LBO and HR scenarios.

#### 1) UNTRUSTED ACCESS

can be understood as the fact that MNO does not trust in the security offered by the non-3GPP access network [6]. Therefore, the traffic must be transported by a secure option from the MNO's point of view. The main component to support the untrusted access network is the Non-3GPP Interworking Function (N3IWF). The fundamental idea of N3IWF, introduced in 3GPP Rel-15 [25], is to act as a gateway for communication between UE and 5GCN. Figure 3 illustrates the connections for the integration between untrusted non-3GPP access networks, especially WLAN or Wi-Fi networks. Moreover, the figure shows encrypted IP Security (IPSec) tunnels, called NWu, that are used to secure transport traffic from untrusted non-3GPP access to the 5G core. In addition, NWu isolates the non-3GPP and 3GPP data. Jungnickel et al. [26] proposed the integration of Li-Fi networks in the same way as Wi-Fi networks, i.e., through N3IWF, which illustrates the flexibility of the untrusted non-3GPP access. The authors also highlight the benefits of integration between non-3GPP access networks and the 5GCN, e.g., the handover support among different networks as an intrinsic feature.

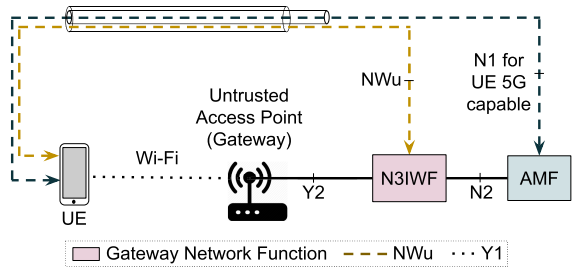

**FIGURE 3.** Connections for the integration between untrusted non-3GPP access network and 5GCN [6].

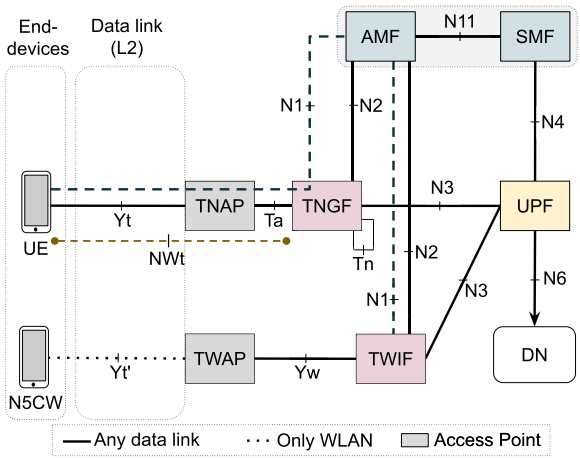

**FIGURE 4.** Connection options to trusted non-3GPP access networks.

### 2) TRUSTED ACCESS

standardized in 3GPP Rel-16 [7], this type of access assumes a different relationship between the non-3GPP access and the 5GCN in comparison with the untrusted scenario. Although the 3GPP standard does not define the trust level [27], we can observe a behavior similar to the 3GPP access. A trusted network indicates that the operator has full control of Trusted Non-3GPP Access Point (TNAP) and the radio link access. Therefore, the encryption is controlled by the operator or there is trust in the security offered by the non-3GPP access network. TNAP enables UEs to access the trusted access network by using non-3GPP wireless or wired access technology. Trusted Non-3GPP Gateway Function (TNGF) exposes N2 and N3 interfaces to allow UE connection to 5GCN over the trusted access network. Trusted Non-3GPP Access Network (TNAN) can be implemented as Trusted WLAN Access Network (TWAN) that, in this case, only supports WLAN [7]. TWAN includes Trusted WLAN Access Point (TWAP) and Trusted WLAN Interworking Function (TWIF) to provide trusted connection to 5GCN for UEs in a WLAN with no 5G capabilities. This type of device is called Non-5G Capable over WLAN (N5CW) and it depends on TWIF for NAS signaling using N1 reference point. In Figure 4, the two trusted options are illustrated: (i) the connection to 5GCN using a generic solution to provide connection to 5GCN through TNAP and TNGF, and (ii) the connection of N5CW devices over WLAN using TWAP and TWIF.

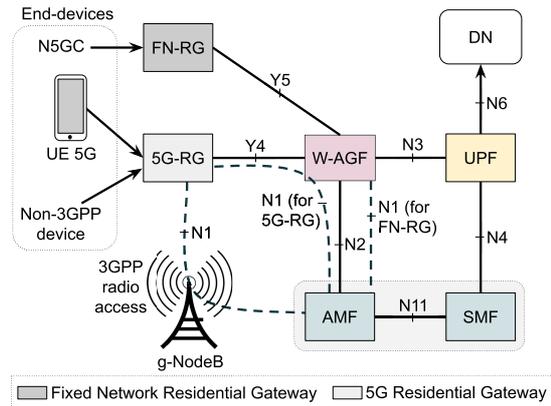

**FIGURE 5.** Possible ways to connect in wireline access networks.

### 3) WIRELINE ACCESS

this type of access was also introduced in Rel-16 by 3GPP [7], mentioning two types of Wireline 5G Access Networks (W-5GAN): (i) Wireline 5G Broadband Access Network (W-5GBAN) and (ii) Wireline 5G Cable Access Network (W-5GCAN), as specified by the Broadband Forum (BBF) [28] and CableLabs® organizations. A gateway function, called Wireline Access Gateway Function (W-AGF), connects these wireline access networks to 5GCN. The 5G Residential Gateway (5G-RG) is a gateway that acts as UE and exchanges NAS signaling with 5GCN. In contrast, Fixed Network Residential Gateway (FN-RG) is a legacy gateway in existing wireline access networks, such as Digital Subscriber Line (DSL) routers, which do not support N1 signaling, i.e., it is not 5G capable. W-AGF provides N2 and N3 interfaces toward 5GCN and may also take over the signaling in N1 interface for end-devices behind FN-RG, i.e., Non-5G Capable (N5GC) devices. Furthermore, W-AGF relays data traffic among residential gateways and UPF [29]. As depicted in Figure 5, 5G-RG can be connected to 5GCN via: (i) W-AGF or (ii) gNodeB, which provides Fixed Wireless Access (FWA). Both accesses can also be adopted simultaneously. A UE, behind 5G-RG, can be connected to the 3GPP radio access network and to the non-3GPP access network through W-AGF simultaneously but employs different instances of N1 interfaces for data traffic.

Table 1 presents a brief comparison between the three types of non-3GPP access networks specified in 3GPP Rel-15 and Rel-16. The trusted and wireline non-3GPP access networks are the options that provide connection to 5GCN from typical devices with no 5G capabilities. In contrast, an untrusted non-3GPP access network considers that UE is capable of supporting NAS signaling. For wireline access, two gateways are defined, 5G-RG and FN-RG, both offering communication to end-devices via W-AGF. Moreover, 5G-RG also offers communication through gNodeB/FWA. N3IWF and TNGF components perform similar functions, but TNGF is considered reliable from the MNO's point of view. In terms of roaming, LBO and HR scenarios are supported in both wireless access networks, except for N5CW devices that only support LBO. The communication is assumed insecure only for untrusted access, which demands additional signaling and secure tunneling. As expected, wireline access does not specify roaming.

**TABLE 1.** Comparison between non-3GPP access networks according to 3GPP Rel-15 and Rel-16.

| Access network type | Traditional end-device | Residential gateway | Type of communication | Main NF | UE 5G capable/NAS signaling support | Roaming support |
|---|---|---|---|---|---|---|
| Untrusted | UE non-3GPP | Access point | Unsecure | N3IWF | Yes | LBO & HR |
| Trusted | UE non-3GPP | TNAP | Secure | TNGF | Yes | LBO & HR |
| Trusted | N5CW | TWAP | Secure | TWIF | No | LBO |
| Wireline | UE non-3GPP | 5G-RG | Secure | W-AGF | Yes | - |
| Wireline | N5GC | FN-RG | Secure | W-AGF | No | - |

## III. UNTRUSTED NON-3GPP ACCESS NETWORK

This section presents an overview of the main related procedures for untrusted non-3GPP access networks. First, we introduce the reference architecture, showing the connections and the relationship of components. Next, the protocols employed for (i) initial access, (ii) IPSec establishment, and (iii) transferring of data packets are discussed to provide a better understanding of registration/authorization and PDU session establishment procedures. Finally, the main procedures to access 5GCN are presented.

### A. REFERENCE ARCHITECTURE

The main component in the untrusted non-3GPP access network is N3IWF, which provides a secure connection to UE for accessing 5GCN via CP/UP functions. This component is used mainly for non-3GPP access, such as Wi-Fi and fixed-line integration into 5GCN. As illustrated in Figure 6, N3IWF connects to AMF via N2 interface in CP. For data traffic in UP, the N3 interface connects N3IWF to UPF. After the authentication and authorization process, non-3GPP devices access 5GCN through a non-3GPP access network to perform NAS signaling via the N1 interface [30]. In this context, the transfer of data packets between UE and DN uses the secure IPSec tunnel between UE and N3IWF. Moreover, the General Packet Radio Services Tunnelling Protocol for User Data (GTP-U) establishes a tunnel between N3IWF and UPF. In summary, N3IWF supports the following main functionalities [7], [31]:

- IPSec tunnel establishment: UE creates NWu reference point using the Extensible Authentication Protocol (EAP) method [32] over Internet Key Exchange Version 2 (IKEv2) [33], and IPSec protocols, e.g., Generic Routing Encapsulation (GRE) [34] and Encapsulation Security Payload (ESP) [35].
- Establishment of IPSec tunnels: (i) for CP, i.e., securing NAS messages, and (ii) for UP, i.e., protecting user plane data traffic.
- Creation of N2 reference point, using Next-Generation Application Protocol (NGAP) and Stream Control

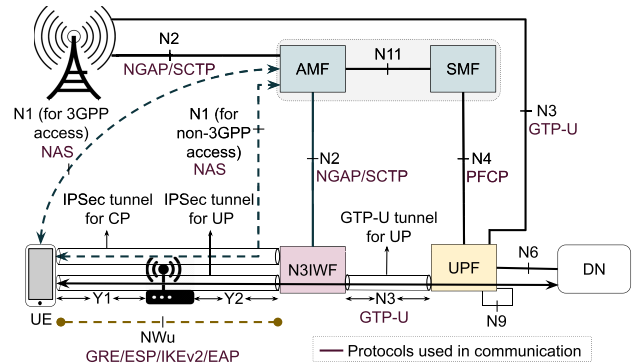

**FIGURE 6.** Architecture for 5GCN with untrusted non-3GPP access network.

Transmission Protocol (SCTP) for CP, and N3 interface, using GTP-U for UP.

- Relaying uplink and downlink for: (i) NAS N1 signaling messages between UE and AMF, and (ii) UP packets between UE and UPF.
- Decapsulation and encapsulation of packets for IPSec and N3 tunneling.
- Handling N2 signaling from SMF relayed by AMF, e.g., related to UP data traffic management session.
- AMF selection, i.e., transparently forwarding messages from UE to AMF.

As illustrated in Figure 6, in untrusted non-3GPP access interworking, both access networks, i.e., 3GPP access via gNodeB and untrusted non-3GPP access via N3IWF, are connected simultaneously. A single AMF serves the UE connected via 3GPP and non-3GPP accesses due to the non-roaming architecture. Multiples N1 instances are created, i.e., one N1 instance over gNodeB and another N1 instance over N3IWF for traffic differentiation.

For both 3GPP and non-3GPP access, the NAS protocol forms the highest stratum of CP between UE and AMF, and SCTP is used to secure the transport of exchanged messages. Both gNodeB and N3IWF connect the N3 interface to UPF, which is encapsulated within GTP-U tunneling. Packet Forwarding Control Protocol (PFCP) [36] is standardized on the N4 interface between CP and UP functions, i.e., to transport creation rules messages from SMF to UPF, being used for the UE traffic classification, queuing, scheduling, and marking/remarking [37].

Table 2 shows the main communication interfaces and the protocols involved in both 3GPP and untrusted non-3GPP

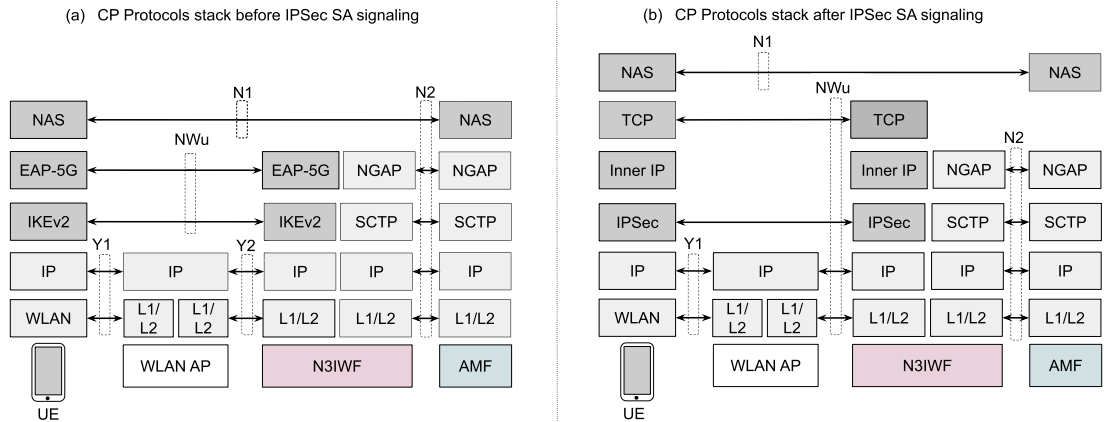

**(a)** CP Protocols stack before IPSec SA signaling

**(b)** CP Protocols stack after IPSec SA signaling access networks. Table lines highlighted in light cyan correspond to 3GPP access. Moreover, this table allows identifying common and different elements in each type of access. NAS protocol is employed to transport the signaling messages for multiples N1 interfaces created for 3GPP access and non-3GPP access. In addition, the source and destination components of the N1 interface are the same due to UE NAS protocol support over untrusted access. NGAP and SCTP are protocols adopted in communication among the components of the N2 interface. For 3GPP access, the source component is gNodeB, whereas, in non-3GPP access, the source of communication is N3IWF. In both cases, the N2 interface is used to forward management session messages to SMF via AMF. N3 interface relays the messages from UE, over gNodeB, or via N3IWF to UPF. GTP-U encapsulates all end-user data over the N3 interface between gNodeB or N3IWF and UPF.

**FIGURE 7.** CP Protocols stacks for NWu connection.

**TABLE 2.** Comparison among communication interfaces, components, and protocols standarized in both accesses networks, 3GPP and untrusted non-3GPP.

| Interface | Source | Destination | Protocols |
|-----------|--------|-------------|-----------|
| N1 | UE | AMF | NAS |
| N1 | UE | AMF | NAS |
| N2 | g-NodeB | AMF | NGAP & SCTP |
| N2 | N3IWF | AMF | NGAP & SCTP |
| N3 | g-NodeB | UPF | GTP-U |
| N3 | N3IWF | UPF | GTP-U |

In addition to the non-roaming case, the untrusted access network supports LBO and HR roaming scenarios. LBO roaming architecture for 5GCN with untrusted non-3GPP access is supported considering two options: (i) N3IWF in the same visited network as 3GPP access, and (ii) N3IWF in a different network from 3GPP access, which is possible due to the independence of the PLMN selection for 3GPP access and non-3GPP access. The main difference between these options is in the selected AMF since the second option UE is served by a different AMF belonging to the same visited network of the N3IWF. In the HR scenario, three options are available:

(i) N3IWF in the same visited network as 3GPP access, (ii) N3IWF in a different visited network than 3GPP access, and (iii) N3IWF in the home network. In options (i) and (iii), UE support in the visited network is provided by the same AMF in 3GPP access and non-3GPP access with multiple N1 instances. However, in (ii), UE is served by different AMF and SMF from 3GPP access.

The authorization process of a UE over untrusted WLAN to access 5GCN and the permission to consume data services are provided by different protocols in CP and UP. It is important to know the protocols involved in the several steps of the UE interaction with 5GCN: (i) for initial access, (ii) before the registration, and (iii) for the establishment of data user transport. Therefore, in the following subsection, we summarize the protocols selected by UE, the WLAN Access Point (AP), N3IWF, and AMF for accessing 5GCN from an untrusted WLAN access network.

### B. CP AND UP PROTOCOLS IN UNTRUSTED WLAN ACCESS
CP and UP protocols standardized for untrusted access vary accordingly with permission obtained by the UE from the core network. Initial access to 5GCN is considered the first contact from UE and is performed before the IPSec signaling due to the absence of a secure tunnel between UE and N3IWF. After registration or IPSec Security Association (SA) establishment, UE can request PDU session establishment to AMF using IPSec tunnel created in NAS signaling.

Figure 7 shows the CP protocol stacks employed by UE, WLAN AP, N3IWF, and AMF before, case (a), and after the IPSec SA signaling, case (b). To provide communication between UE and N3IWF, NAS messages are transferred using EAP-5G/IKEv2. After IPSec SA establishment, UE establishes a Transmission Control Protocol (TCP) connection with N3IWF to transport NAS and session management messages over the Internet Protocol (IP) layer and the IPSec tunnel. AMF adopts the same CP protocol stack to communicate with N3IWF before and after IPSec SA signaling: SCTP and NGAP, via N2 reference point. To communicate with UE, AMF uses NAS protocol to encapsulate messages via the N1 interface.

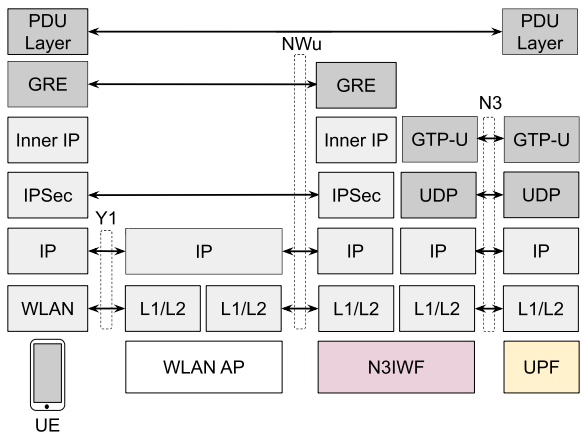

**FIGURE 8.** Protocols in UP for the transferring of PDU sessions.

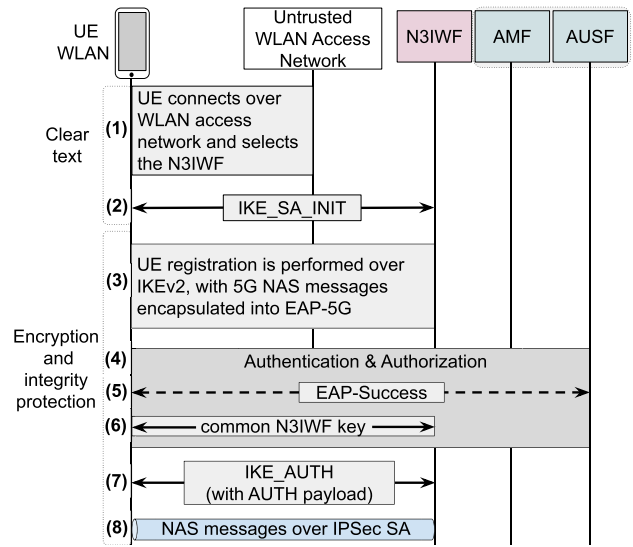

**FIGURE 9.** Signaling IPSec SA establishment over untrusted access.

Figure 8 shows UP protocol stack for transferring data user traffic. PDU layer corresponds to data carried between UE and DN. Before transferring UP traffic, PDU sessions are encapsulated into GRE packets. IPSec tunnel mode is employed for established Child SAs to encrypt and protect the original IP user data packets and the port numbers used for communication. The whole end-user PDU from UE is encapsulated over GTP-U when it arrives in N3IWF before reaching UPF. According to 3GPP TS, which provides the 5GS architecture [38], [39] [40], the UP in 5GS, as it happens in LTE/4G, is still based on GTP-U, and tunneling user-traffic to anchor points in the CN. In addition, GTP-U runs on top of User Datagram Protocol (UDP), allowing many applications, in different ports, with the same IP address.

The protocols in CP permit that a UE connects to 5GCN through registration and authorization procedures to establish an N1 interface for NAS signaling. In contrast, the protocols in UP are used for PDU session establishment. Therefore, we described CP and UP protocols in different stages of communication between UE and 5GCN. In the following, we present an overview of the main procedures related to untrusted non-3GPP access networks.

### C. MAIN PROCEDURES

The access to 5GCN from the untrusted networks involves the following procedures: (i) access network discovery and selection, (ii) registration, authentication, and authorization, and (iii) PDU session establishment. In this subsection, we provide information on how a UE discovers and selects an untrusted non-3GPP network and is authenticated to access 5GCN and authorized to consume data services through UPF.

#### 1) NETWORK ACCESS DISCOVERY AND SELECTION

UE uses the Access Network Discovery and Selection Policy (ANDSP) to discover and prioritize non-3GPP access

networks, e.g., a WLAN Selection Policy (WLANSP) is used for a specified Wi-Fi network and selection of N3IWF in PLMN [41]. UE employs WLANSP for selection and connection to the WLAN access network to obtain an IP address. Non-3GPP access network selection is affected by the type of UE subscription credentials [40]. If UE is roaming, the valid WLANSP in the home or visited network could construct a prioritized list of WLAN access networks.

#### 2) REGISTRATION, AUTHENTICATION, AND AUTHORIZATION

Registration via an untrusted non-3GPP access network adopts a specific vendor EAP method called EAP-5G. This method is used for NAS messages encapsulation over the IKEv2 protocol between UE and N3IWF [39]. First, UE connects to an untrusted access network using an AP, e.g., a UE in a WLAN obtains an IP address. UE selects N3IWF in a PLMN, i.e., perform the procedure in clause 6.3.6.2 of [38] TS if support connectivity with N3IWF and not with LTE/4G or clause 6.3.6.3 if support both, e.g., LTE/4G and 5GCN. Next, UE initiates the procedure to establish IPSec SA, as depicted in Figure 9.

The first messages exchanged between UE and untrusted WLAN access network, in step **(1)**, and between UE and N3IWF, in step **(2)**, are in clear text with no security protection, as shown in Figure 9. For this reason, UE starts the initial procedure with an IKE_SA_INIT, which enables encryption and integrity protection for all subsequent IKEv2 messages. The initial IKE_Auth Request message performed from UE to N3IWF in step **(3)** has no payload, which N3IWF interprets as the request to start an EAP-5G session, i.e., N3IWF responds with an IKE_Auth message containing an EAP-Request/5G-Start packet. This packet informs UE to start an EAP-5G session, i.e., to begin sending NAS messages. Therefore, UE sends a registration request message that includes an EAP-Response/5G-NAS. This IKE_Auth response contains access network parameters that N3IWF uses for the AMF selection process. After selecting AMF, it authenticates UE by invoking AUSF, as shown in step (**4**), which chooses an UDM to obtain authentication data and executes the EAP-AKA or 5G-AKA authentication [42]. For ease of understanding, UDM is not shown in Figure 9. By the end of step (**4**), UE derives the anchor key, NAS, and the N3IWF keys. EAP-5G session is completed at this point, i.e., no further EAP-5G messages are exchanged. After the successful authentication, AUSF sends the Security Anchor Function (SEAF) key to AMF, encapsulated into an EAP-SUCCESS, in step (**5**), in which AMF and UE employ for deriving two keys targeted to enable: (i) NAS security, and (ii) N3IWF security. A derivation process from the anchor key provided from AUSF, UE, and N3IWF has the same standard security key, in step (**6**), i.e., the N3IWF key. The establishment of the IPSec association in step (**7**), known as IPSec SA, is performed using the N3IWF key. In this context, as can be observed in step (**8**), the following NAS messages are transferred over the IPSec SA [39]: finishing registration, authentication, and authorization procedure. From this point on, UE is ready to start the establishment of PDU session for effective data communication, as we will describe in the next subsection.

### 3) PDU SESSION ESTABLISHMENT

Both UE and the network can start the PDU session establishment. UE initiates the PDU session procedure in different cases: (i) for establishing a new session and (ii) for handover proposal. In case (ii), UE initiates a PDU session for handover between 3GPP and non-3GPP networks, e.g., 5G and Wi-Fi, or when the handover is between 5G and EPC. The UE-requested PDU session establishment is fundamental to understanding the procedure used for data communication. Therefore, for didactic reasons, we consider only the UE-requested PDU session establishment for non-roaming and roaming with the LBO scenario. PDU session establishment request is sent by UE to AMF via the IPSec SA for NAS signaling, as defined in step (**1**) of Figure 10. AMF relays the request message according to the roaming context. In general, AMF selects SMF to create a PDU session context. After the PDU session authorization between UE and DN, AMF sends a PDU session request message to N3IWF in step (**2**). N3IWF, based on its policies and Quality of Service (QoS) profiles sent by UE, determines the number of IPSec child SAs created for the association of each received QoS profile in step (**3**). N3IWF sends an IKE message to generate a request to UE for the first IPSec child creation in step (**4**). After the response from UE in step (**5**), additional IPSec SAs can be created, if necessary. After all IPSec SAs being established, PDU session acceptance is sent from N3IWF to UE via the signaling IPSec SA in step (**6**). N3IWF sends to AMF a PDU session response that includes the GTP-U tunnel, as shown in step (**7**). Finally, the QoS flows are transported inside the IPSec child SA previously created by N3IWF requests in step (**8**).

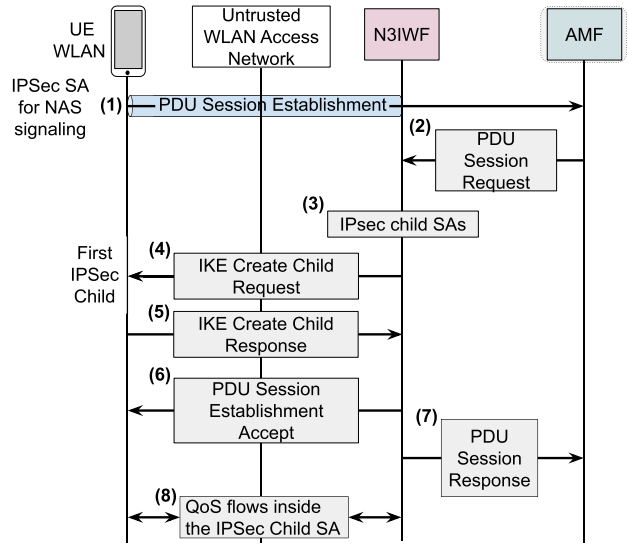

**FIGURE 10.** PDU Session Establishment for untrusted access.

PDU session between UE and N3IWF is encapsulated inside GRE tunnel. UE encapsulates the GRE packets into IP packets with the source address, the inner IP of UE, and the destination IP address associated with the IPSec child SA. When N3IWF receives PDUs from the N3 reference point, N3IWF encapsulates these PDUs into GRE packets and determines the identity of the PDU session for using the correct IPSec child SA. Finally, N3IWF encapsulates GRE packets into IP packets with the source IP address, associated IPSec child SA, and destination inner IP address of the UE. This concludes the description of the untrusted non-3GPP access, which has some commonalities with the trusted non-3GPP access. Thus, in the next section, we present trusted non-3GPP focusing on its unique features.

## IV. TRUSTED NON-3GPP ACCESS NETWORK

This section presents an overview of the main related procedures for trusted non-3GPP access. We employ an organization similar to the previous section, i.e., first, we introduce the reference architecture, and next, the protocols standardized for CP and UP. However, the last part, related to the main procedures, is divided into 5G-capable and non-5G-capable devices (which is not specified in the untrusted non-3GPP access).

### A. REFERENCE ARCHITECTURE

The main component in trusted non-3GPP access networks is TNGF which provides a trusted connection to 5GCN. TNGF has N2 interface to AMF, for CP communication, and N3 interface to UPF for UP functions. The link-layer between UE and TNAP supports EAP encapsulation, as it occurs in untrusted access. The link connection between UE and TNAP can be any link-layer [39], e.g., Point-to-Point Protocol (PPP) [43], Ethernet, IEEE 802.3, Protocol for Carrying Authentication for Network Access (PANA) [44], and IEEE 802.11 [45].

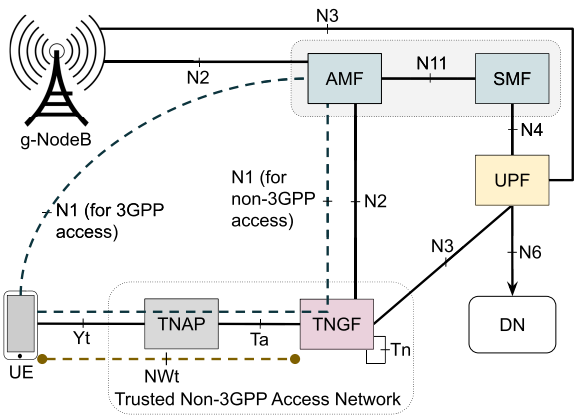

**FIGURE 11.** Non-roaming architecture for 5GCN with trusted access.

From 3GPP point of view, TNGF is a trusted gateway NF, which provides connectivity to UE through a trusted access network, i.e., UE is connected to a trusted access point (TNAP) using non-3GPP wireless or wired access technology. In this context, TNAN is composed of TNAP and TNGF, as illustrated in Figure 11. This figure shows the non-roaming architecture for 5GCN with trusted non-3GPP access, presenting the components of TNAN and the relationship among UE, trusted network, and 5GCN. Similar to untrusted networks, UE using 3GPP access via gNodeB can also connect to trusted non-3GPP access via TNGF.

In summary, the main functionalities supported by TNGF are [38]:

- Handling N2 signaling relayed by AMF.
- Handling N3 interface from UE to UPF.
- AMF selection to transparently forward messages between UE and AMF.
- Relaying of NAS messages between UE and AMF using the secure NWt interface.
- Local mobility support within TNAN and local re-authentication.

UE performs the decision of connecting with an appropriate offer of untrusted or trusted non-3GPP access. If the trusted option is chosen, UE first selects PLMN, and after TNAN, i.e., TNAN employed depends on the selected PLMN [46]. Therefore, in Subsection IV-C, we also describe PLMN selection procedure using a trusted non-3GPP access network.

### B. CP AND UP PROTOCOLS STACK
The registration procedure over trusted non-3GPP access, i.e., before IPSec SA signaling between UE and TNGF, adopts the same vendor-specific EAP-5G procedure of untrusted access. Figure 12 shows the CP protocols stack involved for the NWt connection between UE and TNGF. In case (a), EAP-5G is used to encapsulate NAS messages. The link-layer supports EAP encapsulation, whereas the Authentication, Authorization, and Accounting (AAA) interface is used between TNAP and TNGF. After the NWt connection, in case (b), CP protocol stack becomes the same used in untrusted access: UE establishes a TCP connection with TNGF to transport NAS and session management messages over IP layer and IPSec tunnel. AMF employs the same CP protocol stack to communicate with TNGF before and after NWt connection: SCTP and NGAP. To communicate with UE, AMF uses NAS protocol to encapsulate the messages.

We consider the following features in the trusted scenario due to the focus on WLAN untrusted non-3GPP access:

- Any link-layer can exchange the WLAN access.
- The WLAN AP can be exchanged by a generic AP, i.e., TNAP providing the link-layer to UE.
- TNGF replaces N3IWF.

After IPSec SA signaling, UE establishes a TCP connection with TNGF for transferring all subsequent NAS and session management messages over IP and IPSec layers. IPSec SAs over the NWt interface apply null encryption for both CP and UP to avoid double encryption since the link-layer is trusted. The link-layer is the same in trusted and untrusted networks, e.g., in WLAN, the CP protocol stack standardized for the establishment of UP IPSec child SA is the same for both non-3GPP access networks. For instance, one or more IPSec child SAs are created between N3IWF/TNGF and UE to transfer 5G flows of user data using the IKEv2 protocol.

### C. MAIN PROCEDURES FOR 5G-CAPABLE DEVICES
For 5G-capable devices, access to 5GCN from trusted networks involves the following procedures: (i) network access discovery and selection, (ii) registration and authentication procedure, and (iii) the PDU session establishment.

#### 1) NETWORK ACCESS DISCOVERY AND SELECTION
UE decides to follow the trusted non-3GPP access for connecting to 5GCN based on its own capabilities, e.g., UE supports only trusted access (or related) to discovery non-3GPP access networks. Access Network Query Protocol (ANQP) is standardized to non-3GPP access networks advertise information of PLMNs that the network support. UE can discover the trust relationship for trusted networks over ANQP, i.e., UE sends an ANQP query for request information of the available networks or may be configured with these networks. Each non-3GPP access network may advertise PLMN lists that define the type of supported connectivity. However, if a non-3GPP access network does not support ANQP, Rel-16 of 3GPP [38] does not specify how UE discovers the PLMN lists supported by this non-3GPP access network.

Similar to untrusted non-3GPP access, UE over trusted access is registered, authenticated, and authorized through 5GCN. In trusted access, UE establishes a secure connection to TNGF over NWt interface. After the establishment of IPSec SA with TNGF, UE uses a TCP connection for transporting NAS messages. Furthermore, UE can establish additional child IPSec SAs to transfer user plane traffic.

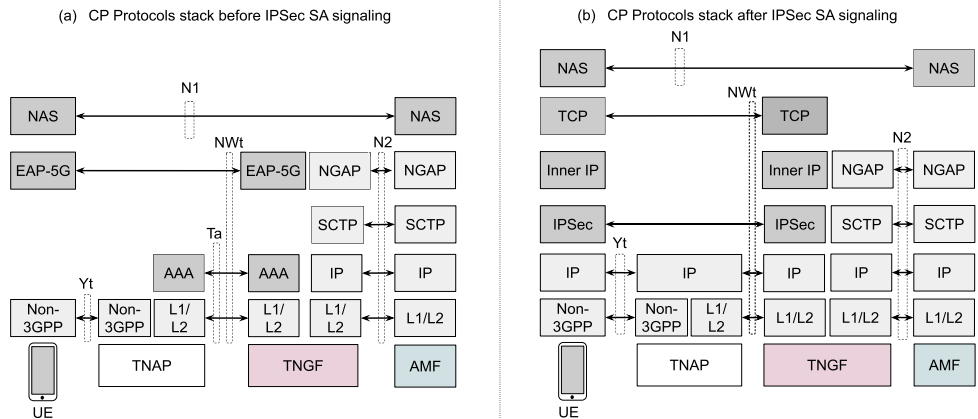

(a) CP Protocols stack before IPSec SA signaling

(b) CP Protocols stack after IPSec SA signaling

**FIGURE 12. CP protocols stacks for the NWt connection.**

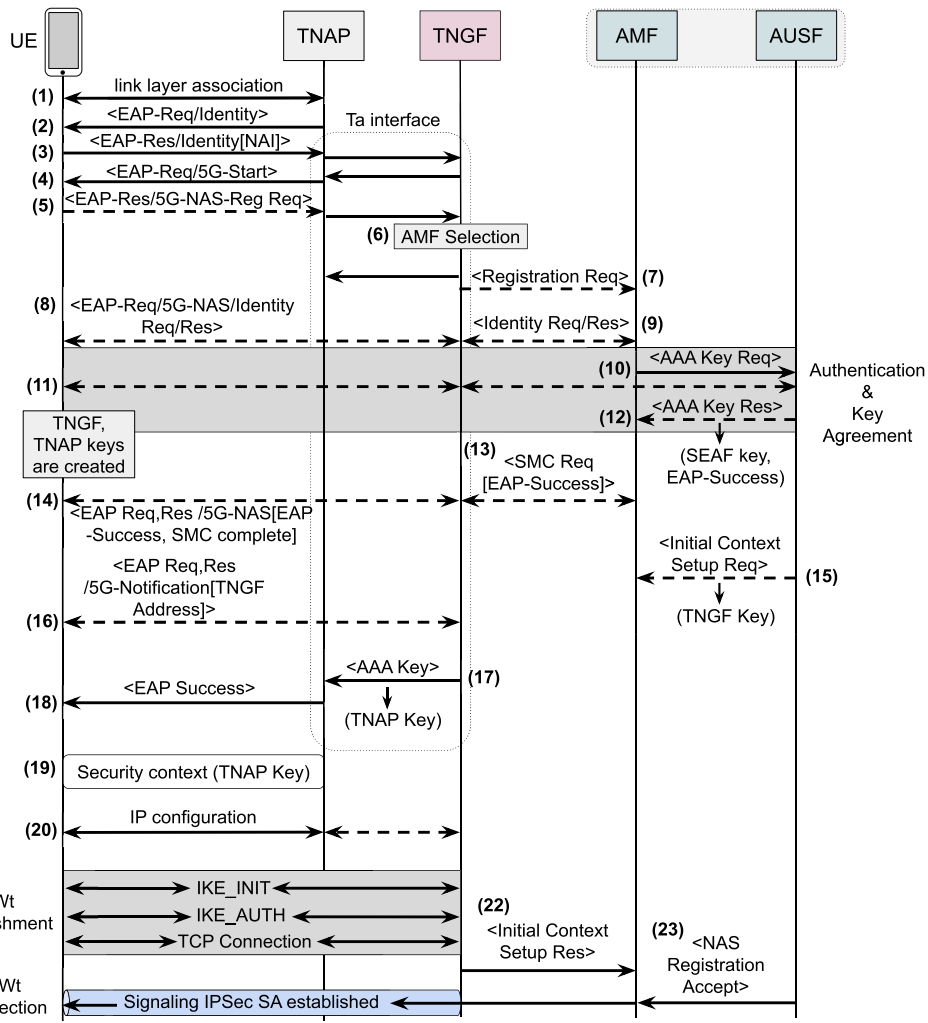

**FIGURE 13. Trusted non-3GPP access registration and authentication.**

## 2) REGISTRATION AND AUTHENTICATION

Figure 13 illustrates trusted non-3GPP access registration and authentication procedure aiming to show:

- How UE derives TNAP/TNGF keys used to protect communication in trusted non-3GPP access networks, i.e., to create IPSec SA with TNGF.

- How TNGF and TNAP receive TNGF and TNAP keys.
- How UE performs the establishment of security context with TNAP using TNAP key.
- How TNGF key is used to perform NWt connection establishment between UE and TNGF.

UE should establish a link-layer association with TNAP in step (**1**), as shown in Figure 13. Any link-layer that supports EAP encapsulation can be used in this case. Link-layer association occurs after access network discovery and selection. For instance, the link-layer association corresponds to an IEEE 802.11 association, whereas Link Control Protocol (LCP) negotiation relates to PPP communications [39]. However, for access networks based on Ethernet technology, the link-layer association may not be required. In step (**2**), TNAP sends to UE an EAP Request for UE Identity, providing a Network Access Identifier (NAI) in EAP Response, in step (**3**), to trigger the EAP-5G session over the Ta interface. UE employs NAI to establish connectivity to a specific PLMN. Similarly, the step (**4**) is an EAP Request indicating that UE starts sending NAS messages.

Messages between UE and TNAP are EAP Requests/ Responses encapsulated into frames, whereas TNAP and TNGF are in AAA format. UE registration request, in step (**5**), is sent to TNGF which performs AMF selection, in step (**6**), to relay the registration message, in step (**7**), to AUSF. Steps (**8**) and (**9**) indicate the messages exchanged between AMF and UE for identification, whereas the steps (**10**)-(**12**) belongs to the authentication and key agreement procedure. In step (**10**), an AAA key request, based on UE identification, is sent to AUSF for an authentication proposal. Step (**11**) represents additional messages for authentication, which are not shown in Figure 13. AUSF sends an AAA key response, in step (**12**), with the security key to UE, e.g., SEAF key may use by UE for the derivation of TNGF/TNAP keys. AMF sends a Security Mode Command (SMC) Request in step (**13**) to UE, which replies with an SMC complete in step (**14**). In general, TNGF key is created by AMF in step (**15**), which is transferred to TNGF for determining the TNAP key. Step (**16**) represents the messages exchanged between UE and TNGF, containing communication information such as TNGF IP address.

In step (**17**), TNGF provides to TNAP the key used for the link-layer security. In this context, step (**18**) shows that the TNGF successfully receives an EAP Response/5G notification packet from UE. Step (**19**) corresponds to the security context established between UE and TNAP with TNAP key. In step (**20**), UE receives an IP configuration from TNAN, i.e., TNAP or TNGF should offer an IP address to UE.

Step (**21**) corresponds to NWt secure connection establishment between UE and TNGF using the IKEv2 protocol. At the end of this process, a TCP connection is established for transfer NAS messages. The process of the NWt establishment holds IKE_INIT and IKE_AUTH exchanged messages between UE and TNGF until the TCP connection establishment. An initial context setup response between TNGF and AMF indicates that NWt connection is successfully established in step (**22**). Finally, AUSF sends to UE a NAS registration acceptance message in step (**23**), informing which AMF forwards to UE. After this point, AMF accepts the registration between TNGF and UE, and all subsequent NAS messages are transported over IPSec SA in step (**24**).

TNAP key is used to protect traffic between UE and TNAP, whereas TNGF key is employed in mutual authentication between UE and TNGF. In this case, TNAP key derivation process depends on non-3GPP access technology. 3GPP specifies the details of how these keys are created [42]. Furthermore, additional details on registration via trusted non-3GPP access networks are described in Technical Report [39] clause 4.12a.2.2. We describe PDU session establishment of UEs that supports NAS in the following.

### 3) PDU SESSION ESTABLISHMENT

UE requests a PDU session establishment using the same procedure specified in III-C3. However, some components of 5GS are different. For example, TNGF substitutes N3IWF and TNAP replaces an untrusted WLAN access network in Figure 10. Additionally, TNGF identities parameters, e.g., IP address is sent from TNGF to AMF, which relays for SMF using N2 interface. Another difference between untrusted and trusted networks concerning PDU session establishment is the creation request of PDU sessions from TNGF to establish a child IPSec SA for QoS flows. TNGF includes additional QoS information that allows UE to determine QoS resources to be reserved in the trusted non-3GPP access network. In contrast to untrusted networks, 5GCN access from UEs is also possible without NAS support, i.e., N5CW devices are supported. In the following subsection, we describe the main procedures for this type of device.

### D. MAIN PROCEDURES FOR NON-5G-CAPABLE DEVICES

5GCN access from devices in WLAN does not support NAS signaling and 3GPP enhances this access in Rel-16, as described in the following.

### 1) INITIAL REGISTRATION

Figure 14 shows the registration procedure for N5CW devices. Similar to initial registration from devices that support NAS signaling, after the access network selection in step (**1**) and data link-layer connection in step (**2**), an EAP-5G procedure is used to verify UE identity, in step (**3**). When TWIF receives NAI from N5CW (step (**4**)), it creates the registration request (step (**5**)) on behalf of the N5CW device, i.e., registration request is not initiated by N5CW. After selecting AMF in step (**6**), TWIF sends a registration request to AMF, authenticating N5CW invoking AUSF, in step (**8**). The registration message uses default values for all N5CW devices. Step (**9**) indicates an EAP-5G authentication procedure between N5CW and AUSF. After successful authentication, AUSF sends a SEAF key to AMF encapsulated into EAP-Success message, as observed in step (**10**). Following, in step (**11**), AMF derives Access Network (AN) key from SEAF key. After this step, a NAS SMC is sent from AMF to TWIF with no security algorithms in step **(12)**. TWIF responds with a SMC complete in step **(13)**. AMF sends an initial context setup request providing AN key to TWIF, in step **(14)**, which derives Pairwise Master Key (PMK) from AN key and sends it to TWAP. PMK is used to secure WLAN communications [47]. PMK is sent encapsulated into EAP-Success to N5CW device in step **(16)**, which derives the WLAN keys, as shown in step **(17)**. Step **(18)** shows the 4-way handshake process used between APs, e.g., TWAP, and N5CW devices, which generate some encryption keys for protecting the communication over the wireless medium. A link (L2) or IP layer (L3) is created between TWAP and TWIF for each N5CW device. Finally, TWIF sends an initial context setup response in step **(20)** to AMF that responds with a registration acceptance message in step **(21)**, finishing the registration procedure. Additional details about the initial registration of N5CW devices are presented in Technical Report [39] clause 4.12b.2. In the following, we describe how TWIF creates PDU session requests and the binding process of L2 or L3 connections created in step **(19)** of Figure 14 to establish the end-to-end communication between N5CW and DN.

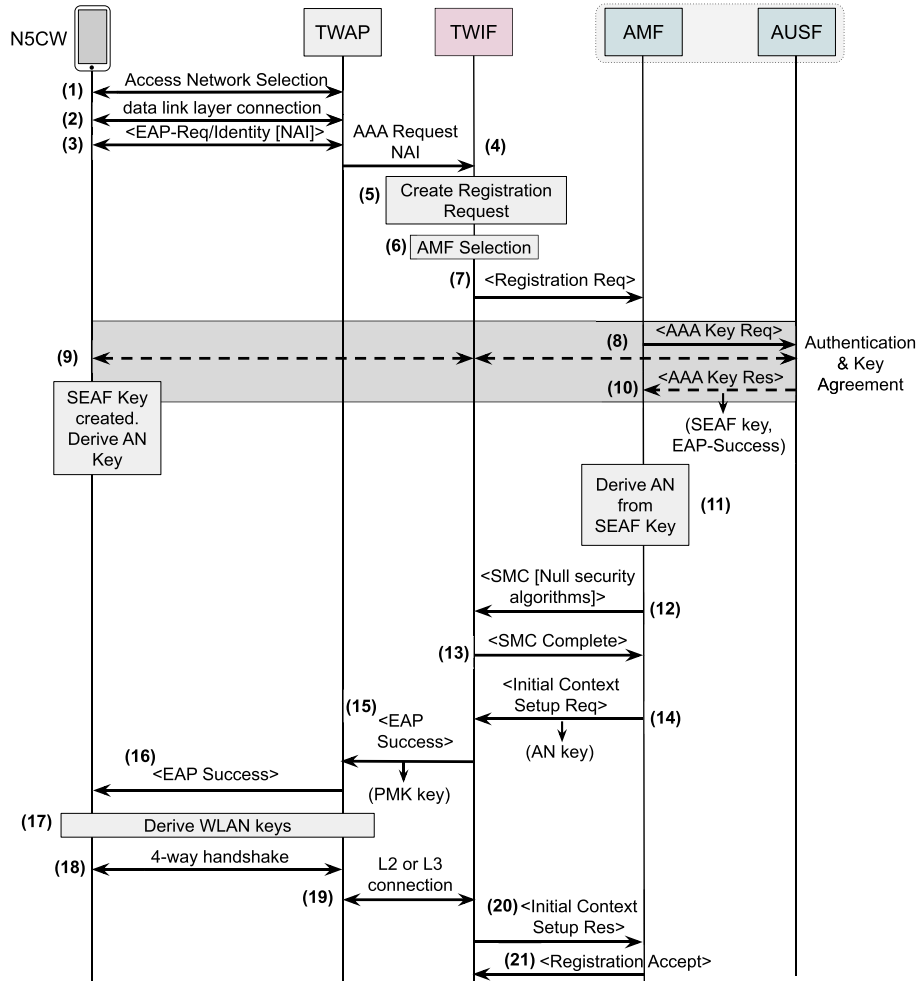

**FIGURE 14.** Initial registration procedure for N5CW devices.

### 2) PDU SESSION ESTABLISHMENT AND UP COMMUNICATION

Figure 15 shows a PDU session establishment for N5CW devices. After successful registration, UE initiates an IP configuration request, in step **(1)**, to TWIF, which triggers the creation of a PDU session establishment request message on behalf of N5CW, as shown in step **(2)**. TWIF fills the message of the PDU session with: (i) default PDU session parameters or (ii) information provided by N5CW devices during the registration procedure. In case (ii), AMF or SMF determines PDU session request message parameters based on N5CW device subscription values. Therefore, in step **(3)**, TWIF sends PDU session establishment Request to AMF. Moreover, if necessary, SMF requests additional information to reserve the appropriate WLAN resources. This session management information is sent from AMF to TWIF, as seen in step **(4)**. Step **(5)** corresponds to WLAN resource reservation, which is an optional process that is outside of the 3GPP scope. After establishing PDU session, N5CW receives an IP address configuration, in step (**6**), which supports PDU session identification. In step (**7**), TWIF sends a PDU session request Ack to AMF. Finally, steps (**8**)-(**9**) represent the UP communication: TWIF binds each L2/L3 connection, created in registration procedure in step (**19**) of Figure 14, to N3 connections created in PDU session establishment, in steps (**4**)-(**7**) of Figure 15.

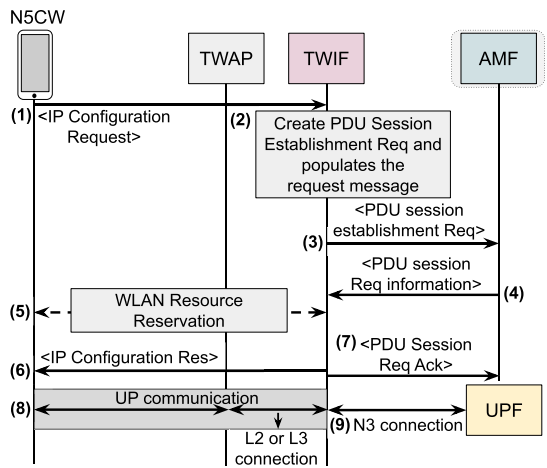

**FIGURE 15.** PDU session and UP communication for N5CW devices.

N3IWF and TNGF have similar CP and UP functionalities. However, Purwita et al. [48] show that there is a key distinction between them when the core network notifies a UE of a trust Wi-Fi network, indicating the importance of the trust concept in non-3GPP access networks. In this context, we argue the relevance of understanding the difference in security aspects between untrusted and trusted networks. This can help MNOs identify requirements for each access network type. We have already shown essential aspects of registration, authorization, authentication, and PDU session establishment of both untrusted and trusted networks. Table 3 compares security-related aspects in untrusted and trusted access networks. In trusted access networks, we considered the two available options for gateways: (i) TNGF and (ii) TWIF.

The common point between non-3GPP access networks is using an EAP-based procedure for the end-devices authentication, as shown in Table 3. Despite being an EAP-based procedure, EAP is not part of the IKEv2 establishment in trusted networks, as it occurs in untrusted networks. The encapsulation between UE and TNAP is performed directly on the access layer, whereas TNAP and TNGF use the Ta interface. In trusted networks, SEAF key derivation process protects communication between UE and AP, performing a link-layer authentication procedure. In this case, there are two options: (i) TNAP key establishes security between UE and TNAP if the trusted network uses TNGF for interworking with 5GCN, or (ii) WLAN keys are used to establish access-specific link-layer security 4-way handshake if TWIF is used for connecting N5CW devices [15], [47]. In both options, SEAF key derivation performs link-layer authentication between UE and AP. On the other hand, untrusted networks do not use SEAF key for derivation access network key. Security between UE and untrusted AP can be established with any key, including no security. In this case, untrusted networks do not perform link-layer authentication between UE and AP. In untrusted networks and trusted networks with the TNGF option, IPSec SAs apply an appropriate key derived from the SEAF key. N3IWF key is used between UE and N3IWF, and TNGF key is employed between UE and TNGF. There is a trust relationship in trusted networks, i.e., the link-layer encryption is trusted, allowing IPSec encryption, if applicable, to negotiate null security options avoiding double encryption.

From 5G, there is also the intention of promoting wireless and wireline convergence, allowing 5GCN access from non-3GPP networks over several media and technologies. In the next section, we describe 3GPP specification for non-3GPP wireline access.

## V. WIRELINE NON-3GPP ACCESS NETWORK
In this section, initially, we introduce reference architectures for W-5GAN, in which CableLabs® and BBF contribute to the development of 5G-RG and FN-RG solutions. Next, we present a specific integration architecture developed by CableLabs® and BBF scenarios focusing on components, connections, and communication interfaces. Finally, CP and UP protocol stacks are explained in the context of the registration procedure and PDU session establishment for both 5G-RG and FN-RG.

### A. REFERENCE ARCHITECTURES FOR W-5GAN
The complete convergence in a 5G network is achieved when a single 5GCN provides access to different networks, i.e., wireless or wireline access. In this context, BBF and CableLabs® are responsible for promoting wired access network technologies to 5GCN. BBF introduces 5G Broadband Residential Gateway (5G-BRG) as a 5G-RG and Fixed Network Broadband RG (FN-BRG) as an FN-RG. CableLabs® specifies 5G Cable Residential Gateway (5G-CRG) as a 5G-RG and Fixed Network Cable RG (FN-CRG) as an FN-RG. As described in Subsection II-B, independent of the specific BBF or CableLabs® solution to support Residential Gateway (RG) for 5GS, W-AGF is employed for connectivity with 5GCN, i.e., W-AGF provides N2 and N3 interfaces to related functions with CP and UP, respectively.

From 3GPP point of view, W-AGF belongs to a W-5GAN, which is considered a layer of interworking capabilities between the wireline network and 5GCN. BBF and CableLabs® have their own implementations of W-5GAN. In the case of modems using DOCSIS, W-5GCAN is the implementation of W-5GAN. However, instead of broadband services compliant to BBF, 5G convergence is described introducing the Access Gateway Function (AGF) and an alternative access mechanism.

**TABLE 3.** Comparison among security aspects between untrusted and trusted networks.

| Type of non-3GPP network | EAP based authentication procedure | Security key between UE and AP | Link-layer authentication procedure | SEAF key derivation for access network key | SEAF key derivation for IPSec key | Security key for IPSec SA establishment | IPSec Encryption |
|---|---|---|---|---|---|---|---|
| Untrusted | Yes | Any key | No | No | Yes | N3IWF Key | Yes |
| Trusted (TNGF) | Yes | TNAP key | Yes | Yes | Yes | TNGF Key | Null |
| Trusted (TWIF) | Yes | WLAN keys | Yes | Yes | - | - | - |

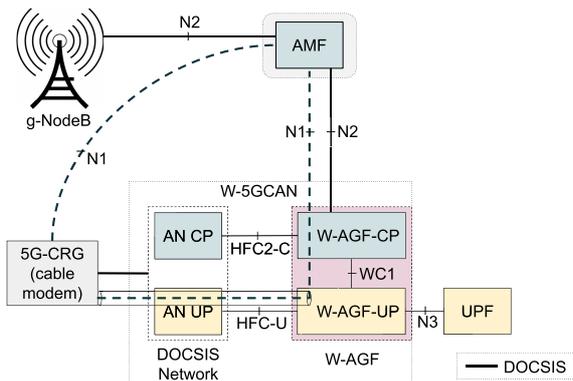

**FIGURE 16.** CableLabs integration architecture for Cable 5G RG.

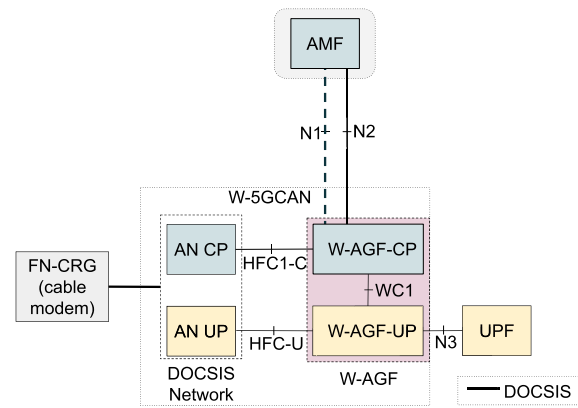

**FIGURE 17.** CableLabs integration architecture for Fixed Cable RG.

Figure 16 shows a CableLabs® integration architecture for 5G-CRG that offers hybrid access to 5GCN. 5G-CRG connects to AMF via gNodeB using the N1 interface and via wireline access using a DOCSIS network. From CableLabs® point of view, W-5GCAN includes DOCSIS network and W-AGF. AN CP interfaces to Wireline access Control Plane (W-CP) via HFC2-C uses the existing high-speed Wide Area Network (WAN) interface and should be standardized to achieve interoperability among multiple vendors [49].

The integration model for FN-CRG in the 5G converged network architecture is depicted in Figure 17. The management of the N1 interface by the W-AGF component, more specifically by W-CP used for NAS signaling between 5G-RG and W-AGF, is notably different than the approach used in the previous architecture (shown in Figure 16). W-AGF communicates with AMF over the N1 interface and with SMF via the N2 interface. Differently from 5G-CRG, interface HFC1-C is used between AN CP and W-AGF-CP.

According to BBF specifications [50], [51], the introduction of AGF allows the convergence of 5G-RG and FN-RG. Based on Trick [46], Figure 18 shows all possible convergence scenarios developed by BBF. The six bold numbers on the left of the figure represent these scenarios, whereas the five numbers on the right represent the sets of related interfaces. In the following, we briefly describe all these numbers.

1) **FWA with 5G-RG** has a 3GPP radio interface to connect to gNodeB, which supports N1 interface and offers direct communication with 5GCN.

2) **Hybrid access combining gNodeB and fixed accesses.** 5G-RG supports N1, N2, and N3 interfaces, and a wireline interface with an N1 interface to 5GCN.

N2 and N3 interfaces are provided by an intermediate AGF, whereas 3GPP access is provided by gNodeB.

3) **Direct mode with 5G-RG** has a wireline interface and no 3GPP interface. N2 and N3 interfaces are provided by AGF, as it occurs in scenario (2).

4) **Adaptive mode with FN-RG** represents RG without supporting an N1 interface. In this case, UEs behind FN-RG are N5GC, and AGF provides all 3GPP interfaces, e.g., N1, N2, and N3 to 5GCN.

5) **Interworking with FN-RG** uses intermediate functions, e.g., Broadband Network Gateway (BNG) and Fixed Mobile Interworking Function (FMIF), as alternative access for legacy devices behind FN-RG.

6) **FN-RG in coexistence** is a direct by-pass of 5GCN via BNG that is directly connected to the DN. BNG is the only interface adaptation necessary to connect FN-RG to 5GCN.

Scenarios (5) and (6) are outside of the 3GPP scope. 3GPP standard is aware of 5G-RGs and FN-RGs, while CableLabs® and BBF organizations provide specific solutions for W-5GAN in accordance with 3GPP point of view. In the following subsection, we overview of the CP and UP protocol stacks involving 5G-RG, FN-RG, and 5GS entities to support the wireline non-3GPP access network.

### B. CP AND UP PROTOCOL STACKS
CP protocol stack between 5G-RG and AMF adopts NAS for signaling messages. W-AGF is connected to 5G-RG via the Y4 interface (shown in Figure 5) and employs a generic component W-CP to transport functions and NAS signaling [52]. The specific protocols of W-CP stack depend on the type of 5G-RG. W-CP stack used for 5G-BRG is defined in BBF specifications [51], whereas W-CP stack used for 5G-CRG is defined in CableLabs [49]. The communication between W-AGF and AMF using the N2 reference point is based on SCTP and NGAP. CP protocol stack between FN-RG and 5GCN is based on another component that depends on the type of legacy RG. Legacy Wireless access Control Plane (L-W-CP) protocol stack is used in W-5GBAN, between FN-BRG and W-AGF, and is also defined in BBF specifications [51]. In W-5GCAN, L-W-CP protocol stack between FN-CRG and W-AGF is defined in CableLabs® [49] and BBF [53]. Protocols standardized to transport PDU sessions (between 5G-RG and W-AGF over Y4 interface) use Wireline access User Plane (W-UP) function, which is considered part of W-AGF, similar to W-CP [52]. The definition of W-UP also relies on BBF and CableLabs®.

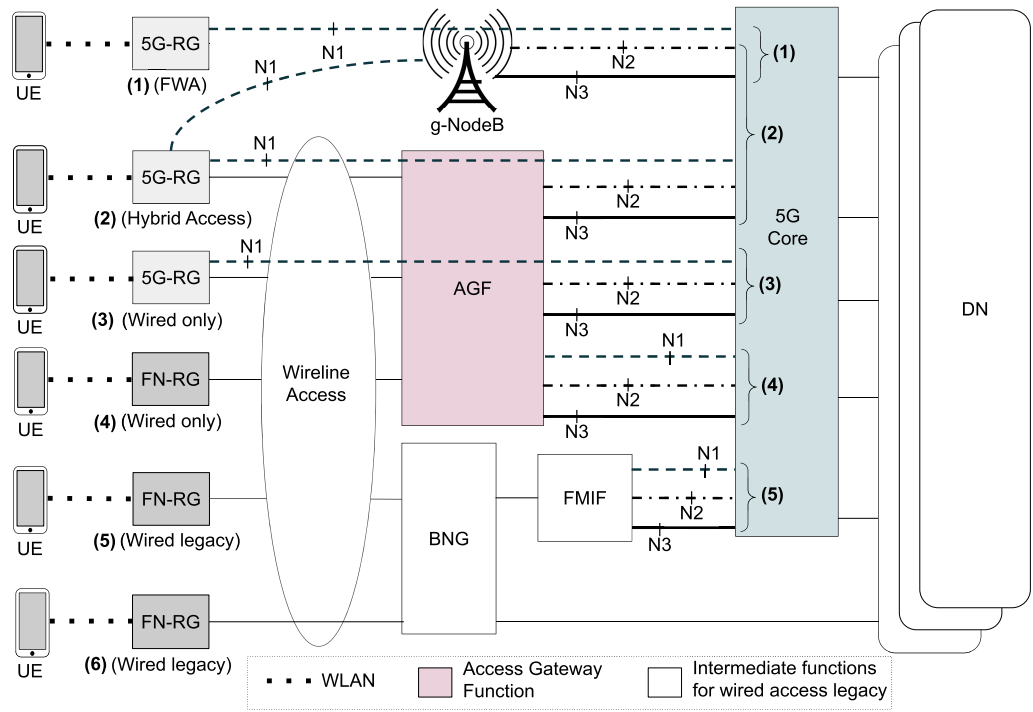

**FIGURE 18.** Access network scenarios for Fixed Mobile Convergence with 5GCN developed by BBF [46].

**TABLE 4.** Components of W-AGF applied in the context of W-5GBAN and W-5GCAN and their specifications.

| Components | Protocol stack used for | W-5GBAN | | W-5GCAN | |
|---|---|---|---|---|---|
| | | 5G-BRG | FN-BRG | 5G-CRG | FN-CRG |
| W-CP | Control Plane | [51] | - | [49] | - |
| L-W-CP | Control Plane | - | [51], [53] | - | [49] |
| W-UP | User Plane | [51] | - | [49] | - |
| L-W-UP | User Plane | - | [51], [53] | - | [49] |

Considering 5G-RGs, the W-UP protocol of W-AGF communicates with 5G-BRG or 5G-CRG. W-UP protocol stack definition between 5G-BRG and W-AGF is described in BBF specifications [51], whereas between 5G-CRG and W-AGF is found in CableLabs® [49]. In the context of FN-RGs, Legacy

Wireless access User Plane (L-W-UP) specification between FN-BRG and W-AGF is defined in BBF specifications [51]. For W-5GCAN, the communication between FN-CRG and W-AGF using the L-W-UP protocol stack is specified in CableLabs® [49]. To facilitate the identification of CP and UP protocol stacks used between RGs in 5GS, we provide Table 4 that summarizes the components belonging to W-AGF in both configurations, W-5GBAN and W-5GCAN. This table also references where to find the details specified by BBF and CableLabs®.

For CableLabs® [49], 5G-CRG may appear as both 5G UE and an FN-CRG. On the one hand, operators may decide to integrate both into a single element. On the other hand, other operators will deploy 5G-CRGs where the 5G UE is implemented separately from the FN-RG. The modular implementation approach permits independent provisioning and management systems. According to [50], an RG will incorporate at least one embedded WAN interface, routing, bridging, a basic or enhanced firewall, one or multiple LAN interfaces and home networking functions that can be deployed as a consumer self-installable device.

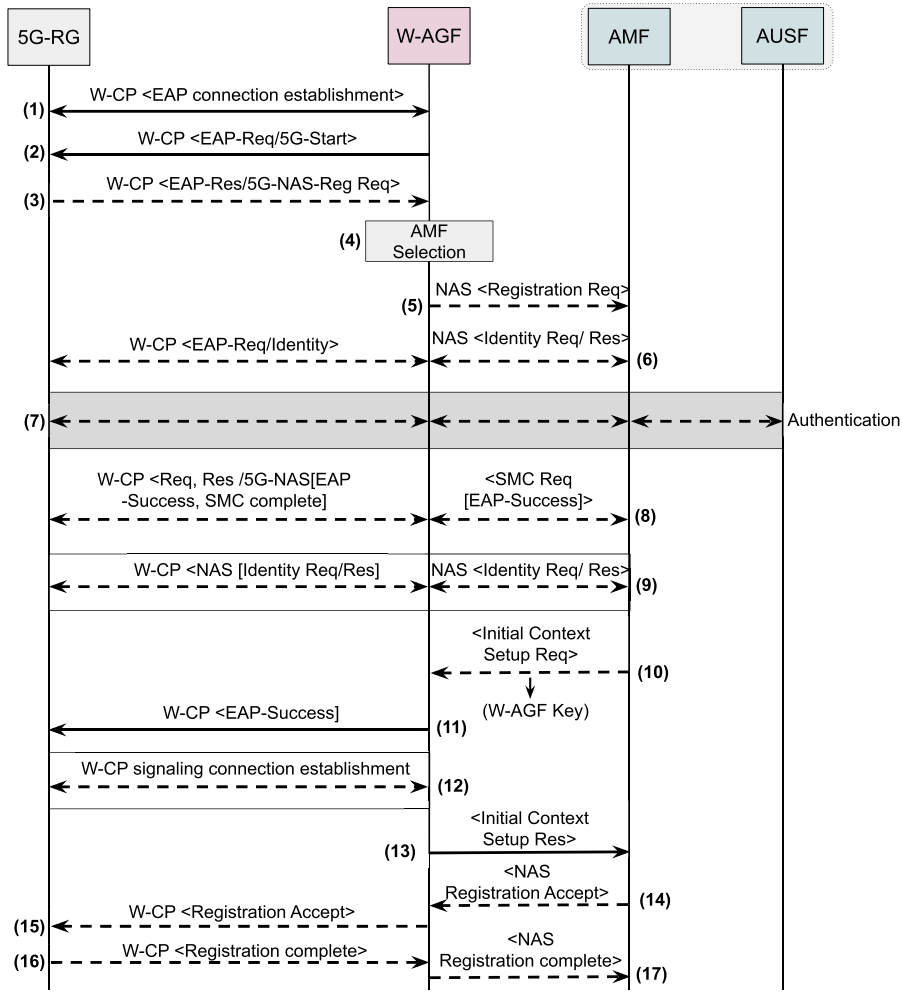

**FIGURE 19.** 5G-RG registration procedure.

W-AGF holds L-W-CP and L-W-UP to provide communication with the legacy RGs developed by BBF and CableLabs®. Specific details about W-AGF are defined by these third-party organizations, as indicated in Table 4. In the following, we provide an overview of the main procedures related to wireline non-3GPP access from 3GPP point of view.

### C. MAIN PROCEDURES

The procedure for registering 5G-RG and FN-RG in the 5GCN relies on W-AGF to intermediate the communication process. As we present in the following, this component is involved in the two main procedures: registration process (for both 5G-RG and FN-RG) and PDU session establishment via W-5GAN (for both W-5GBAN and W-5GCAN). In summary:

1) **5G-RG Registration**: procedure used by 5G-RG to communicate with W-AGF from 3GPP point of view.

2) **FN-RG Registration**: procedure used by legacy devices to communicate with W-AGF from 3GPP point of view.

3) **PDU Session Establishment via W-5GAN**: PDU sessions performed by 5G-RG and FN-RG to reach DN using wireline access.

### 1) 5G-RG REGISTRATION

First, 5G-RG establishes a connection with W-AGF in step **(1)**, as shown in Figure 19. In this case, despite CableLabs® and BBF deciding specific procedures for wireline connection, for 3GPP [52], EAP-5G is used to encapsulate messages between 5G-RG and W-AGF. In step **(2)**, W-AGF sends an EAP request using the W-CP connection established in the previous step. 5G-RG interprets this message as a request to start sending NAS messages, shown in step **(3)**. After W-AGF performs AMF selection in step **(4)**, NAS registration request from 5G-RG is forwarded to AMF, as described in step **(5)**. AMF requests (and receives) the identity of W-AGF in step **(6)**, while step **(7)** represents the authentication of 5G-RG. After AMF invoking AUSF, which has details suppressed in Figure 19, AMF sends information about 5G-RG to AUSF for authentication proposal. If AUSF successfully authenticates 5G-RG, AMF sends an SMC Request, in step (**8**), to 5G-RG, which replies with an SMC complete. AMF starts an initial context setup request with the W-AGF key received from AUSF in step (**8**). Step (**9**) is a conditional step performed if a new AMF requests the identification messages exchange from 5G-RG again. Step (**10**) represents the sending of W-AGF key from AMF to W-AGF that forwards an EAP success using the W-CP connection in step (**11**), and so finishes EAP-5G session. The conditional step (**12**) represents the establishment of a W-CP signaling connection between 5G-RG and W-AGF. After this signaling connection, if applicable, W-AGF sends an initial context setup response to AMF, in step (**13**), indicating the success of the context initiated by AMF. Finally, AMF sends NAS registration acceptance message, in step (**14**), which is forwarded by W-AGF to 5G-RG (step (**15**)). When 5G-RG receives this message of success in the registration procedure, it sends a registration complete, in step (**16**), to AMF, which is forwarded back by W-AGF, as shown in step (**17**). A detailed description of the 5G-RG registration via W-5GAN is specified in the Technical Report "Wireless and wireline convergence access support" [52], clause 7.2.1.1. In the following subsection, we provide the registration procedure of legacy devices, i.e., for FN-RGs.

### 2) FN-RG REGISTRATION

FN-RG establishes a connection to W-AGF via a data-link layer, in step (**1**), as shown in Figure 20. After AMF selection in step (**2**), using AN parameters and local policy, W-AGF performs NAS registration request on behalf of FN-RG to the 5GCN, as described in step (**3**). In general, if FN-RG connects to W-AGF for the first time, Subscription Concealed Identifier (SUCI) from FN-RG is used for W-AGF in the registration request forwarded to AMF. SUCI is built by W-AGF accordingly to the type of access network, i.e., BBF or CableLabs®. As it occurs in the 5G-RG registration procedure after AMF invoking AUSF, AMF sends information about FN-RG to AUSF for authentication proposal. AUSF performs UDM selection (not illustrated in the figure) to perform a map between SUCI to Subscription Permanent Identifier (SUPI) for FN-RG. Step (**4**) represents the successful authentication. In step (**5**), W-AGF sends an SMC Request to W-AGF, which replies with an SMC complete in step (**6**). Initial Setup Context Request message is sent from AMF to W-AGF in step (**7**), which replies creating a context in an Initial Setup Context Response message, as shown in step (**8**). Finally, NAS registration acceptance is sent from AMF to W-AGF in step (**9**), which finishes the procedure by sending NAS registration complete in step (**10**). A detailed description of the FN-RG registration via W-5GAN is also specified in the Technical Report "Wireless and wireline convergence access support" [52], clause 7.2.1.3.

### 3) PDU SESSION ESTABLISHMENT VIA W-5GAN

As we describe in the following, PDU session establishment can be summarized in a few steps for both 5G-RG and FN-RG. 5G-RG sends PDU session establishment to W-AGF using the W-CP signaling connection, and W-AGF forwards it to AMF [52]. After sending the PDU session resource setup request to AMF, W-AGF also determines W-UP resources of the PDU session for the 5G-RG as defined by BBF and CableLabs® specific implementations. The complete procedure description of the PDU session requested for 5G-RG (Technical Report [52] clause 7.3.1.1) is based on 3GPP access, using additional parameters of W-AGF identity.

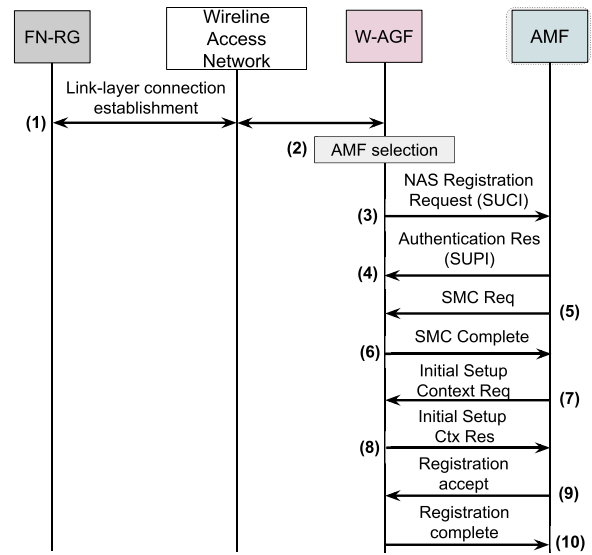

**FIGURE 20.** FN-RG registration procedure.

FN-RG does not send the PDU session establishment request to AMF, differently from 5G-RG scenario. This step is performed by W-AGF due to the absence of NAS capabilities from FN-RG. Therefore, W-AGF performs this procedure on behalf of FN-RG. The trigger to initiate a PDU session establishment also depends on BBF and CableLabs® specific implementations. For 3GPP [52], PDU session establishment for FN-RG is performed on 3GPP access. According to 3GPP, the full description of the FN-RG PDU session establishment via W-5GAN is specified in [52] clause 7.3.4.

Until this point in the tutorial, we have presented a consistent overview of all options available for non-3GPP access. This background is very important for infrastructure providers/operators, but it is also useful for service providers or verticals interested in or depending on non-3GPP technologies. To bring the concepts closer to practical use, we present in the next section a basic proof-of-concept that covers registration, authorization, and PDU session establishment on an untrusted non-3GPP access network.

## VI. PRACTICE: UNTRUSTED NON-3GPP ACCESS WI-FI USE CASE

This section presents a 5GCN access over an untrusted Wi-Fi non-3GPP access network. Initially, we describe the experimental environment, including the software tools and configuration of components belonging to 5GS. Next, we provide a basic performance evaluation involving the main procedures related to registration and authorization and PDU session establishment. Our experimental environment and all experiments presented in this section can be replicated following detailed instructions publicly available in GitHub repository.[1] We also made publicly available a generic UE software[2] that can be extended for any untrusted non-3GPP access technology. Our UE software and some other contributions are part of *my5G* [54], an open-source initiative focused on making 5G systems modular and easy to deploy and operate, turning 5G accessible also to non-experts.

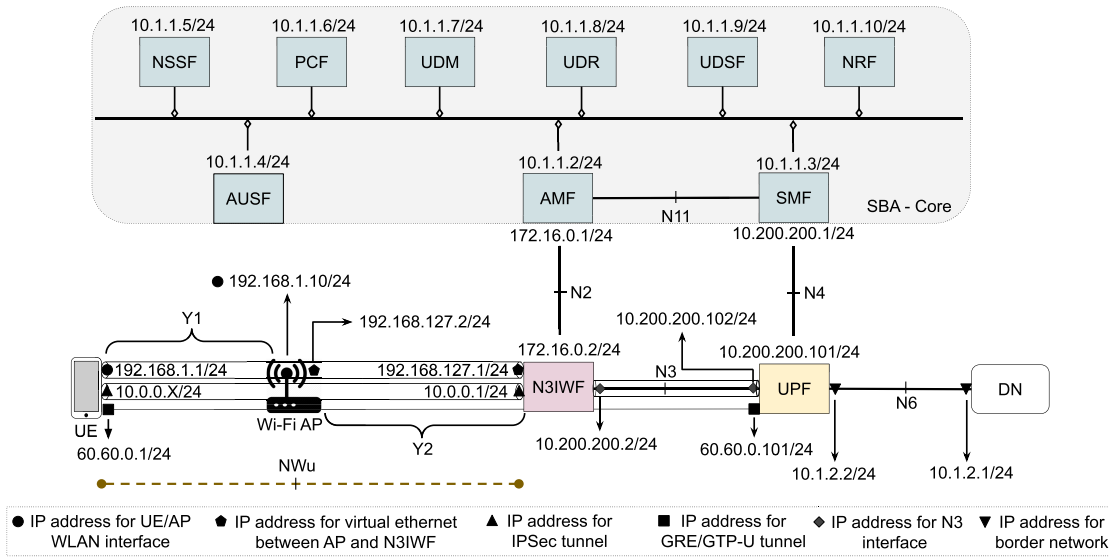

**FIGURE 21.** Architecture of our untrusted non-3GPP access PoC using Wi-Fi network.

### A. EXPERIMENTAL ENVIRONMENT

Figure 21 illustrates the architecture of our PoC of an untrusted non-3GPP access network in which Wi-Fi technology is employed. All software employed in the PoC is open-source publicly available, starting from the Operating System (OS), which is Linux Ubuntu 18.04. Our UE software and the 5G core are based on free5GC project,[3] which implements a basic but fully functional Standalone 5G core according to 3GPP Rel-15. Our UE software implements all necessary functionalities for testing registration, authorization, and PDU session establishment, including secure connections and signaling.

To make the PoC flexible and easy to replicate, we employ *mac80211_hwsim* [55] tool to simulate IEEE 802.11 radios of UE and Wi-Fi AP. We create four different namespaces (`default`, `UEns`, `APns`, and `UPFns`) to isolate components of 5GS and properly mimic a real-world scenario. CP NFs and N3IWF are in `default` namespace, whereas UPF is in `UPFns`. UE simulated radio belongs to `UEns` and AP simulated radio belongs to `APns`.

We use *dnsmasq* [56] as a Dynamic Host Configuration Protocol (DHCP) server for automatically assigning IP addresses to UE. We use *hostapd* tool [57] for enabling a *mac80211_hwsim* (virtual) interface to act as an AP, and *wpa_supplicant* [58] software to implement key negotiation and IEEE 802.11 authentication and association. Moreover, we create virtual Ethernet devices and bridges to connect the components in Figure 21. To closer represent a real-world scenario and also make a clear separation of the components, we adopt the IP addressing shown in the figure. This configuration proved to be flexible for several experiments and easy to debug. The whole environment can be replicated in the host OS or in a virtual machine, including one in a cloud computing infrastructure. Finally, we employ *Wireshark* [59] tool to capture packets and analyze protocols.

NFs in CP, which interact with each other using an SBA paradigm, are in 10.1.1.0/24 subnet. The secure interface between UE and N3IWF, i.e., NWu reference point, is in the 10.0.0.0/24 subnet. In this case, the IP address of end-point IPSec tunnel in N3IWF-side is always the same, 10.0.0.1/24, whereas the IP address of IPSec tunnel in UE-side varies randomly within the available IP range 10.0.0.2/24 - 10.0.0.254/24. Finally, PDU sessions between UE and UPF are transferred using 60.60.0.0/24 subnet: (i) UE uses 60.60.0.1/24 IP address in GRE tunnel between UE and N3IWF, and (ii) UPF uses 60.60.0.101/24 IP address in GTP-U tunnel between N3IWF and UPF.

### B. PERFORMANCE EVALUATION

The evaluation focuses on controlling the overhead of untrusted non-3GPP access while performing some important procedures and during data transfer in UP. Only the critical correspond to 30 trials of each procedure. Both procedures consume a notably long time to perform compared to commonly advertised 5G latency values. However, those procedures are rare, mainly IPSec SA signaling, which is performed when UE is turned on and connects to 5GCN in-home or visited network (in roaming), for example. A PDU session establishment is necessary when UE connects the DN (commonly the Internet), i.e., it is more frequent than registration and authorization procedure, but it is still not frequent and performed before the effective data transfer.

---

[1]https://github.com/LABORA-INF-UFG/paper-MACAK-2022

[2]https://github.com/my5G/my5G-non3GPP-access

[3]https://www.free5gc.org/

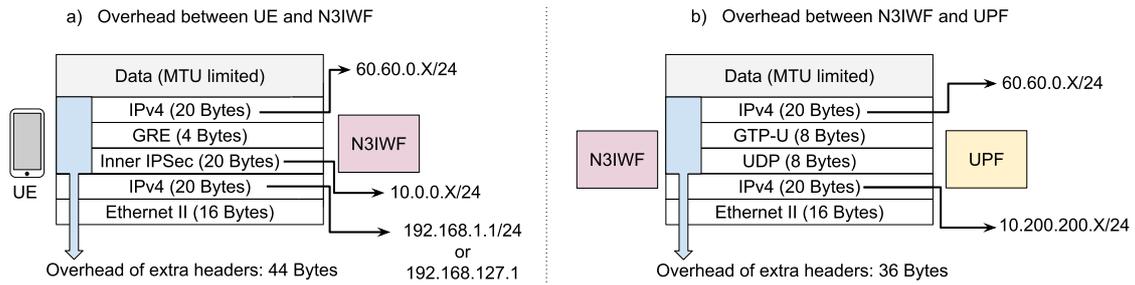

**FIGURE 22.** Protocols overhead of a PDU session.

**TABLE 5.** Time statistics about some procedures.

| Procedure | Average (s) | Standard deviation (s) | Confidence interval of 95% | |
|---|---|---|---|---|
| | | | Lower limit (s) | Upper limit (s) |
| IPSec SA Signaling | 0.93 | 0.41 | 0.78 | 1.07 |
| PDU Session Establishment | 0.22 | 0.04 | 0.21 | 0.23 |

**TABLE 6.** IPSec SA signaling messages.

| ID | Message | Involved Components | Size (Bytes) |
|---|---|---|---|
| 1 | IKE_SA Init Request (0) | UE → N3IWF | 644 |
| 2 | IKE_SA Init Response (0) | N3IWF → UE | 644 |
| 3 | IKE_Auth Request (1) | UE → N3IWF | 216 |
| 4 | IKE_Auth Response (1) | N3IWF → UE | 1448 |
| 5 | IKE_Auth Request (2) | UE → N3IWF | 200 |
| 6 | InitialUEMessage | N3IWF → AMF | 128 |
| 7 | Authentication Request (1) | AMF → N3IWF | 148 |
| 8 | IKE_Auth Response (2) | N3IWF → UE | 168 |
| 9 | IKE_Auth Request (3) | UE → N3IWF | 152 |
| 10 | Authentication Response (1) | N3IWF → AMF | 140 |
| 11 | Security Mode Command | AMF → N3IWF | 124 |
| 12 | IKE_Auth Response (3) | N3IWF → UE | 152 |
| 13 | IKE_Auth Request (4) | UE → N3IWF | 184 |
| 14 | UplinkNASTransport | N3IWF → AMF | 168 |
| 15 | InitialContextSetupRequest | AMF → N3IWF | 188 |
| 16 | IKE_Auth Response (4) | N3IWF → UE | 120 |
| 17 | IKE_Auth Request (5) | UE → N3IWF | 136 |
| 18 | IKE_Auth Response (5) | N3IWF → UE | 296 |
| 19 | InitialContextSetupResponse | N3IWF → AMF | 100 |
| 20 | DownlinkNASTransport | AMF → N3IWF | 160 |
| | **Exchanged messages between UE and N3IWF** | | **4360** |
| | **Internal exchanged messages between NFs** | | **1156** |
| | | **Total** | **5516** |

**TABLE 7.** PDU session establishment messages.

| ID | Message | Involved Components | Size (Bytes) |
|---|---|---|---|
| 1 | PFCPSessionEstablishmentReq | SMF → UPF | 271 |
| 2 | PFCPSessionEstablishmentResp | UPF → SMF | 107 |
| 3 | PDUSessionResourceSetupReq | AMF → N3IWF | 271 |
| 4 | Echo Request | N3IWF → UPF | 58 |
| 5 | Echo Response | UPF → N3IWF | 58 |
| 6 | Create_Child_SA Req | N3IWF → UE | 488 |
| 7 | Create_Child_SA Res | UE → N3IWF | 456 |
| 8 | PDUSessionResourceSetupRes | N3IWF → AMF | 120 |
| | **Exchanged messages between UE and N3IWF** | | **944** |
| | **Internal exchanged messages between NFs** | | **885** |
| | | **Total** | **1829** |

Table 6 and Table 7 list messages exchanged in the previous procedures, i.e., IPSec SA signaling and PDU session establishment, respectively. Messages between UE and N3IWF are in light cyan rows and messages between pairs of 5G core functions are in white rows. As shown in the tables, while the traffic volume is low, several messages are involved, mainly in the IPSec SA signaling procedure. Most messages are small (less than 300 Bytes), but several are related to security issues that traditionally demand some processing time.

Figure 22 illustrates protocols overhead of a PDU session established between UE and N3IWF, and between N3IWF and UPF. As previously described, a secure tunnel is created between UE and N3IWF using IPSec and GRE, which implies an extra 44 Bytes of headers. GTP-U tunnel between N3IWF and UPF adds 36 Bytes of extra headers. In the companion public repository, we present detailed information about the headers of all these protocols and also describe how to capture and analyze packets from 5G CP and UP.

path of each procedure is presented, i.e., we do not show results involving errors or optional steps. We analyze control messages exchange and data packet headers. We do not add artificial delays in our experimental environment, thus time-related results are dominated by end-systems (e.g, UE, N3IWF, AMF) processing.

Table 5 summarizes time statistics related to two procedures: IPSec SA signaling, i.e., initial registration and authorization, and PDU session establishment. These results

## VII. FINAL REMARKS

This article presents a comprehensive tutorial on 5GCN access via non-3GPP access networks. Initially, we introduced 5GS and types of non-3GPP access networks specified in Rel-15 and Rel-16 of 3GPP. After, we described registration, authentication, and authorization of end-devices that, in the sequence, consume data services through PDU sessions establishment with the 5GCN. For untrusted and trusted networks, we also discussed network access discovery and selection, considering detailed protocols in different stages of communication between UE and 5GCN, i.e., (i) before the registration, (ii) after IPSec SA signaling, and (iii) for PDU session establishment. Moreover, we compared the security characteristics of these networks, such as authentication and encryption procedures.

Concerning wireless wireline convergence in 5GS, we presented wireline non-3GPP access network and authentication and authorization procedures of residential gateways, describing the possibilities for wired connection to 5GS. Finally, to illustrate the background provided by the tutorial, we presented a PoC of untrusted non-3GPP access using Wi-Fi. We evaluated some performance metrics, such as the time consumed in some important procedures, the number of messages, and the overhead of the UP in this access network. The information provided by this tutorial is important in the context of non-3GPP access for 5G but also B5G/6G networks.

Rel-16 has made significant contributions by developing trusted non-3GPP access network option mainly due to access from N5CW devices. Thus, it makes possible the 5GCN access from UEs that do not have a 3GPP interface to support NAS signaling messages. However, the wireless wireline convergence is limited for providing service continuity between 5G, legacy RG, and AP supported by wireless, optical, copper, and fiber media. Multi-access capabilities are improved in Rel-16, allowing the maximization of data rates, increased reliability, and improved user experience.

Release 17 (Rel-17) should extend this initiative of integration and convergence to individual services at home, providing unified access-agnostic solutions and policy-driven service experience. Furthermore, Rel-17 will extend traffic steering capabilities among mobile and fixed accesses, enhancing the convergence between non-3GPP access technologies and CN. We argue that this fact could help the research community to identify potential topics and contributions related to 5GCN access via non-3GPP networks that could strongly impact on Rel-17 and beyond, and also on novel protocol design and architecture approaches for future networks.

### ACKNOWLEDGMENT
The authors want to thanks CNPq and CAPES.

## LIST OF ACRONYMS

**3GPP** Third Generation Partnership Project
**5G** Fifth Generation

**5G-BRG** 5G Broadband Residential Gateway
**5G-CRG** 5G Cable Residential Gateway
**5GCN** 5G Core Network
**5G-NR** 5G New Radio
**5G-RG** 5G Residential Gateway
**5GS** 5G System
**AMF** Access and Mobility Function
**AGF** Access Gateway Function
**AN** Access Network
**ANDSP** Access Network Discovery and Selection Policy
**ANQP** Access Network Query Protocol
**AP** Access Point
**AF** Application Function
**AAA** Authentication, Authorization, and Accounting
**AUSF** Authentication Server Function
**BBF** Broadband Forum
**BNG** Broadband Network Gateway
**CP** Control Plane
**CUPS** Control and User Plane Separation
**CN** Core Network
**DN** Data Network
**DSL** Digital Subscriber Line
**DHCP** Dynamic Host Configuration Protocol
**ESP** Encapsulation Security Payload
**eMBB** Enhanced Mobile Broadband
**EAP** Extensible Authentication Protocol
**EPC** Evolved Packet Core
**FMIF** Fixed Mobile Interworking Function
**FN-BRG** Fixed Network Broadband RG
**FN-CRG** Fixed Network Cable RG
**FN-RG** Fixed Network Residential Gateway
**FWA** Fixed Wireless Access
**GRE** Generic Routing Encapsulation
**HTTP** Hypertext Transfer Protocol
**HPLMN** Home Public Land Mobile Network
**HR** Home-Routed
**IKEv2** Internet Key Exchange Version 2
**IP** Internet Protocol
**IPSec** IP Security
**L-W-CP** Legacy Wireless access Control Plane
**L-W-UP** Legacy Wireless access User Plane
**LCP** Link Control Protocol
**Li-Fi** Light Fidelity
**LBO** Local Breakout
**LTE** Long Term Evolution
**ML** Machine Learning
**mMTC** Massive Machine Type Communications
**MNO** Mobile Network Operator
**NAI** Network Access Identifier
**NEF** Network Exposure Function
**NSSF** Network Exposure Function
**NF** Network Function
**NRF** Network Slice Selection Function
**NGAP** Next-Generation Application Protocol
**gNodeB** Next-Generation Node B
**NG-RAN** Next-Generation Radio Access Network

**N3IWF** Non-3GPP Interworking Function
**N5GC** Non-5G Capable
**N5CW** Non-5G Capable over WLAN
**NAS** Non Access Stratum
**NPN** Non-Public Network
**OS** Operating System
**PDU** Packet Data Unit
**PFCP** Packet Forwarding Control Protocol
**PMK** Pairwise Master Key
**Point-to-Point Protocol over Ethernet (PPPoE)** Point-to-Point Protocol over Ethernet
**PCF** Policy Control Function
**PLMN** Public Land Mobile Network
**PoC** Proof-of-Concept
**QoS** Quality of Service
**RAN** Radio Access Network
**Rel-15** Release 15
**Rel-16** Release 16
**Rel-17** Release 17
**RG** Residential Gateway
**SEAF** Security Anchor Function
**SA** Security Association
**SBA** Service-Based Architecture
**SBI** Service-Based Interface
**SEPP** Security Edge Protection Proxy
**SMF** Session Management Function
**SMC** Security Mode Command
**SDN** Software-Defined Networking
**SCTP** Stream Control Transmission Protocol
**SUCI** Subscription Concealed Identifier
**SUPI** Subscription Permanent Identifier
**TLS** Transport Layer Security
**TCP** Transmission Control Protocol
**TNAN** Trusted Non-3GPP Access Network
**TNAP** Trusted Non-3GPP Access Point
**TNGF** Trusted Non-3GPP Gateway Function
**TWAN** Trusted WLAN Access Network
**TWAP** Trusted WLAN Access Point
**TWIF** Trusted WLAN Interworking Function
**URLLC** Ultra-Reliable Low Latency Communications
**UDM** Unified Data Management
**UDP** User Datagram Protocol
**UE** User Equipment
**UP** User Plane
**UPF** User Plane Function
**VPLMN** Visited Public Land Mobile Network
**WAN** Wide Area Network
**W-5GAN** Wireline 5G Access Network
**W-5GBAN** Wireline 5G Broadband Access Network
**W-5GCAN** Wireline 5G Cable Access Network
**W-CP** Wireline access Control Plane
**W-AGF** Wireline Access Gateway Function
**W-UP** Wireline access User Plane
**WLAN** Wireless Local Area Network
**WLANSP** WLAN Selection Policy

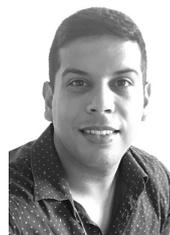

**MARIO TEIXEIRA LEMES** received the bachelor's degree in computing engineering from the Pontifícia Universidade Católica de Goiás (PUC-GO), in 2011, and the M.Sc. degree in computer science from the Universidade Federal de Goiás (UFG), Brazil, in 2014, where he is currently pursuing the Ph.D. degree in computer science. He has been a Professor with the Instituto Federal de Goiás, since 2014. His research interests include networking area, such as next-generation networks (5G/6G), software-defined networking, network function virtualization, wireless networks, and intent-based networks.

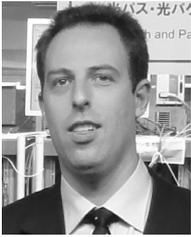

**ANTONIO MARCOS ALBERTI** received the M.Sc. and Ph.D. degrees in electrical engineering from Campinas State University (Unicamp), Campinas, São Paulo, Brazil, in 1998 and 2003, respectively. He has been the Head of the ICT Laboratory and an Associate Professor and a Researcher with the Instituto Nacional de Telecomunicações, Natal, Brazil, since 2004. In 2012, he was a Visiting Researcher at the Future Internet Department, ETRI, South Korea. Since 2008, he has been a designing and implementing a future internet architecture called NovaGenesis. He also works with 6G.

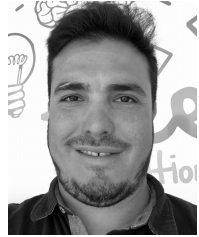

**ANTONIO CARLOS DE OLIVEIRA JÚNIOR** (Member, IEEE) received the B.S. degree in computing and the M.S. degree in electrical engineering from the Federal University of Uberlândia, Brazil, in 2000 and 2003, respectively, and the Ph.D. degree in computer science (Doctoral Programme in computer science) from Minho, Aveiro, and Porto Universities (MAP-i), Portugal, in 2014. He has been a Professor with the Institute of Informatics (INF), Federal University of Goiás (UFG), since 2004. From 2007 to 2008, he was a Researcher of Telecommunications Institute (IT-Porto). From 2008 to 2010, he was a Researcher of INESC-TEC, Porto. From 2011 to 2014, he was a Researcher of the SITILabs and COPELABS (Association for the Research and Development of Cognition and People-centric Computing), Lisbon. From February 2019 and February 2020, he spent his sabbatical at Fraunhofer Portugal AICOS which actually is an Invited Senior Scientist. He is currently a member of Laboratory (Computer Networks and Distributed Systems) mainly involved in networking area, such as the Internet of Things, smart city, wireless networks, energy efficiency, intelligent networking, UAVs, MEC, and 5G/B5G NGN.

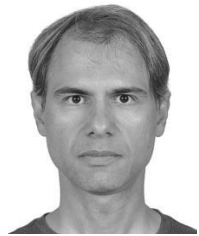

**KLEBER VIEIRA CARDOSO** (Member, IEEE) received the degree in computer science from the Universidade Federal de Goiás, in 1997, and the M.Sc. and Ph.D. degrees in electrical engineering from COPPE—Universidade Federal do Rio de Janeiro, in 2002 and 2009, respectively. He is currently an Associate Professor with the Institute of Informatics, Universidade Federal de Goiás (UFG), where he has been a Professor and a Researcher, since 2009. In 2015, he spent his sabbatical at Virginia Tech (USA). In 2020, he was at the Inria Saclay Research Centre (France). He has participated in some international research projects (including two from joint calls BR-EU) and coordinated several national-sponsored research and development projects. His research interests include wireless networks, software-defined networks, virtualization, resource allocation, and performance evaluation.

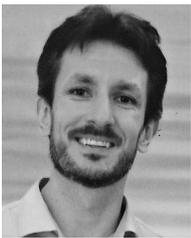

**CRISTIANO BONATO BOTH** is currently an Associate Professor of the applied computing graduate program with the University of Vale do Rio dos Sinos (UNISINOS), Brazil. He has coordinated research projects funded by H2020 EU-Brazil, CNPq, FAPERGS, and RNP. His research interests include wireless networks, next generation networks, softwarization and virtualization technologies for telecommunication networks, and SDN-like solutions for the Internet of Things. He is participating in several Technical Programme and Organizing Committees for major international conferences.

● ● ●